\documentclass[a4paper,10pt, floatfix,showpacs, preprintnumbers, superscriptaddress, nofootinbib, onecolumn]{revtex4}
\usepackage{amssymb,amsmath,amsfonts,amsbsy,graphicx,rotating,bbold}
\usepackage{color,microtype}
\usepackage{breqn}
\usepackage{yfonts}
\usepackage[cal=boondoxo]{mathalfa}
\usepackage{tabularx}
\usepackage{eucal}
 \usepackage{caption}
\usepackage{epsfig}
\usepackage{url}
\usepackage{geometry}
\usepackage{float}
\usepackage{xcolor}
\usepackage{lipsum}
\usepackage{textcomp}
\usepackage{siunitx}
\usepackage{makecell}
\usepackage{lipsum}
\usepackage{hyperref}

\hypersetup{
colorlinks=true,
linkcolor=blue,
filecolor=red,   
citecolor= red,
urlcolor=magenta}

\graphicspath{{Images/}}

\usepackage{subcaption}
\definecolor{vividviolet}{rgb}{0.62, 0.0, 1.0}
\definecolor{amaranth}{rgb}{0.9, 0.17, 0.31}
\definecolor{palatinateblue}{rgb}{0.15, 0.23, 0.89}
\definecolor{brightpink}{rgb}{1.0, 0.0, 0.5}
\definecolor{cornflowerblue}{rgb}{0.39, 0.58, 0.93}
\definecolor{deepcarminepink}{rgb}{0.94, 0.19, 0.22}
\definecolor{radicalred}{rgb}{1.0, 0.21, 0.37}

\def\beq{\begin{equation}}
\def\eeq{\end{equation}}

\begin{document}

\title{Cosmological evolution with quadratic gravity and nonideal fluids}

\author{Saikat Chakraborty$^\dagger$, Daniele Gregoris$^\ddagger$
\\
{\it $^\dagger$Center for Gravitation and Cosmology, College of Physical Science and Technology, Yangzhou University
\\ 
180 Siwangting Road, Yangzhou City, Jiangsu Province  225002, China
\\
$^\dagger$School of Aeronautics and Astronautics, Shanghai Jiao Tong University, Shanghai 200240, China
\\
$^\dagger$International Center for Cosmology, Charusat University, Anand 388421, Gujrat, India
\\
$^\ddagger$School of Science, Jiangsu University of Science and Technology, Zhenjiang 212003, China }
\\
$^\dagger$saikatnilch@gmail.com, $^\dagger$snilch@yzu.edu.cn, $^\ddagger$danielegregoris@libero.it}


\begin{abstract}
Some cosmological models based on the gravitational theory $f(R)=R+\zeta R^2$, and on  fluids obeying to the  equations of state of Redlich-Kwong,  Berthelot, and Dieterici are proposed for describing  smooth transitions between different cosmic epochs. A dynamical system analysis reveals that these models contain fixed points which correspond to an  inflationary, a radiation dominated and a late-time accelerating epoch, and a nonsingular bouncing solution, the latter being an asymptotic fixed point of the compactified phase space. The infinity of the compactified phase space is interpreted as a region in which the non-ideal behaviors of the previously mentioned cosmic fluids are suppressed. Physical constraints on the adopted dimensionless variables are derived  by demanding the theory to be free from ghost and tachyonic instabilities, and a novel cosmological interpretation of such variables is  proposed through a cosmographic analysis. The different effects of the equation of state parameters on the number of equilibrium solutions and on their stability nature are clarified. Some generic properties of these models, which are not sensitive to the particular fluid considered, are identified, while  differences  are critically examined by showing that the Redlich-Kwong scenario admits a second radiation-dominated epoch and a Big Rip Singularity.
\end{abstract}

\maketitle

\section{Introduction}

Despite being a challenging task, the search for a unified cosmological theory accounting for the entire known evolution history of the Universe, or at least providing a smooth transition between two different cosmic epochs, has been attempted both through single fluid approaches and by proposing modifications of the gravity sector beyond general relativity \cite{chap1,bamba}. In the former case a certain single cosmic fluid is adopted to describe two different epochs in the limits of high and low energy, while the latter framework postulates some curvature modifications to the Lagrangian which are dominant at a certain cosmic epoch but dilutes at others. For example, Born-Infeld-like theories can lead to an effective description of the cosmic matter interpolating between dark matter and dark energy dominated epochs as a consequence of the Friedman equations, in terms of the Chaplygin Gas \cite{chap3,chap4} or of the Anton-Schmidt fluid \cite{chap5,chap6}. Other thermodynamically-motivated fluid models like the Dieterici \cite{chap7} or the Shan-Chen \cite{chap8} can as well exhibit a phase transition from a decelerating to an accelerating phase of the universe; the former from a matter-dominated epoch to a dark energy epoch, and the latter from an early radiation-dominated epoch to a dark energy epoch. The Shan-Chen model can also be used for describing the exponential expansion occurring during the inflationary epoch with the advantage of exhibiting a graceful exit mechanism, but for a different choice of the free parameters entering its equations of state than in the former analysis \cite{chap9}.  

On the other hand, extended gravity theories in which a certain curvature invariant is added to, or used to replace, the Ricci scalar inside the Einstein-Hilbert Lagrangian can provide as well an evolution between different cosmic epochs as a consequence of the modified field equations themselves \cite{rev1,rev2,DeFelice:2010aj,res1,resa,resb,noj}. In spite of the correspondence between modified gravity theories and non-ideal fluid pictures (i.e. whose pressure and energy density are connected via $P=w(\rho)\rho$) \cite{capoequiv}, the former have the advantage of not violating some of  the energy conditions which instead are broken when exotic fluids with negative pressure are adopted, and they preserve causality which would be lost when the adiabatic speed of sound squared becomes negative. 

In this paper, we will merge the fluid and the modified gravity approaches and propose some cosmological models in which the gravity sector is accounted for by a Lagrangian of the type $f(R)=R+\zeta R^2$, while the matter content is assumed to obey to some non-ideal equations of state with a well-established thermodynamical foundation known under the names of Redlich-Kwong, (Modified) Berthelot, and Dieterici fluid separately. The former assumption will allow us to account for the early-time dynamics, while the latter for the present-day epoch. Both these two models have been investigated separately in a number of literature works \cite{hysto1,hysto2,hysto3,hysto4,hysto5,hysto6,hysto7,capo,epjc2020,mcvittie}. Here, we will obtain a cosmological dynamics with a rich variety of different behaviors like a non-singular bounce,  two de Sitter-like epochs (thanks to the non-linear equation of state of the cosmic fluid in which $w(\rho)$ is not a constant), possibly two radiation-dominated epochs, and possibly a phantom regime (the latter only in the Redlich-Kwong scenario). The comparison between three different realizations of the equation of state parameter function $w(\rho)$ (for example which can be either always regular or admitting singularities, which can blow up or not at small or high energy densities, etc...) will give us the opportunity of enlightening which of our findings hold only when a particular fluid modeling is considered, and which instead seem to be a general characteristic of the cosmological dynamics. We must mention here that previously there have been some attempts to unify early and late time cosmology under certain forms of $f(R)$ gravity \cite{Nojiri:2003ft,Cognola:2007zu,Nojiri:2007cq}. However, it is worthwhile to remark also that the modifications utilized in those works are completely ad-hoc, lacking any motivation from the field theory point of view. The only modifications to the Einstein-Hilbert Lagrangian with some field theoretical motivations are the quadratic gravity theories. It has been known for some time that gravity Lagrangian containing additional quadratic curvature invariant terms are renormalizable \cite{Utiyama,Stelle}. Therefore in this work we do not intend to go beyond quadratic modifications. In particular we consider only the simplest  case, namely, an $R^2$ correction term, along with fluids having a well-defined thermodynamic foundation.

We will tackle the technical difficulties arising in a fourth-order gravity theory like this one by adopting the set of dimensionless variables constructed in \cite{carloni} which allows to cast the dynamical equations into a system of autonomous first-order equations suited for a dynamical system analysis. Such technique constitutes a powerful mathematical tool for describing the qualitative evolution of the the cosmological model under investigation not only in modified gravity \cite{carloni,revmd,Chakraborty:2018bxh,carlonicapo,fourthorder,Goheer:2007wx,comptexp,comptlif,qual1,qual2,jib}, but also in multi-interacting fluid models \cite{epjc2020,intref1,intref2,int1,int2,int3,sergeim,mcg3,biswas}, and in exact or perturbed anisotropic and inhomogeneous cosmological models \cite{buch1,buch2,buch3,buch4,buch5,buch6,buch7}, just to mention a few examples. However, we will also propose a novel cosmologically transparent interpretation for those variables which was still lacking in the literature by deriving the physical restrictions they should obey to for avoiding tachyonic and ghost instabilities and connecting them to the cosmographic parameters, such as the deceleration, jerk and snap parameters which can be astrophysically constrained. Remarkably, we will show that such physical restrictions still allow the existence of a region in the phase space in which the energy density of the matter field is equal to the energy density of the curvature, which may be relevant for addressing the coincidence problem.  Furthermore, our choice of variables will be useful also for showing that certain regions of the phase space are free from any of the five known types of cosmological singularities without the need of using the dominant energy balance formalism \cite{bal}. Moreover, after compactifying the phase space we will show that the region at infinity does not have only a geometrical meaning but it is such that the cosmic fluid equation of state reduce to the ideal behavior $P \propto \rho$ in which the interactions between the fluid constituents are suppressed.

One of the most severe shortcomings of the standard cosmological modeling is the Hubble tension, which is the discrepancy between the large and small scale estimates of the Hubble constant from supernova and cosmic microwave background data. Assuming that these predictions are not affected by any systematics, as  to gravitational lensing effects on the cosmic microwave background angular spectrum \cite{alens1} or to calibration and reddening issues for supernovae \cite{alens2,alens3,alens4}, an appropriate theoretical framework should be constructed for taming it. Several different proposals have been formulated, but none of them  still seem  fully satisfactory. For example the presence of a Proca field would reduce the Hubble tension \cite{proca1}, but there are no laboratory evidences of massive electrodynamic effects, and furthermore gauge invariance is lost in this theory \cite{jackson}. Also, interactions between dark energy and dark matter may alleviate the tension \cite{intb1}, but thermodynamical considerations based on the Le Chatelier-Braun principle suggests that dark energy should decay into dark matter \cite{intb2} while the fact that the structure formation era should precede the accelerating phase would require otherwise \cite{intb3}.  Our present work is intended as a rigorous dynamical study of a unified cosmic history model, combining two important frameworks one each from the study of early and late-time universe. Although we do not address the issue of $H_0$ tension here, an interesting scope for further investigation would be whether a unified cosmic history model, like the one we presented here, can provide an alternative to introducing ad-hoc interactions in the dark sector when it comes to alleviating the $H_0$ tension. Indeed this is not the first time that modified gravity and other ingredients are merged together. For example, anisotropic models in which the Copernican principle is relaxed have already been considered in Einstein-Aether gravity \cite{aether1} also with  a coupling to a scalar field \cite{buch2},  in braneworld cosmologies \cite{aether3},  or in $f(R)$ gravity \cite{aether4}, just to cite a few examples.    On the other hand, for  a  recent  phenomenological proposal 
which may tame some observational challenges invoking two free parameters and requiring only a modification of the gravity sector in terms of a torsional Lagrangian   see \cite{saridakis}.  

Our paper is organized as follows:  in Sect. \ref{secII} we will review the field equations of the class of models we want to analyze and exhibit the equations of state of the cosmic fluids we are adopting mentioning their basic features, and we will as well introduce a formalism in which both curvature and matter effects are combined into an effective picture. Sect. \ref{main} constitutes the main part of our work: in \ref{condder} we will recast the equations governing the dynamics of our models as a system of autonomous first order equations in terms of a set of dimensionless variables on which we will also derive appropriate physical restrictions;  in \ref{qualsec} we will identify the cosmologically meaningful equilibrium solutions, explain for which ranges of the matter equation of state parameters they can arise pointing out possible bifurcations among them for particular types of matter contents, and report their stability showing that radiation-dominated, de Sitter-like and power law cosmologies can arise; in \ref{sectcomp} we will compactify the phase space and perform the analysis at infinity showing that a nonsinglar bounce occurs; in \ref{y=2zman}, \ref{omegamani} and \ref{R=1sub} we will investigate the dynamics in the invariant submanifolds both numerically by plotting the trajectories in the phase spaces, by deriving analytically their stability, and by finding analytical results for the phase orbits in some specific cases; in \ref{cosmop} we will relate the dimensionless variables we have adopted to the deceleration, jerk and snap cosmographic parameters which can be astrophysically measured. Then, in Sect. \ref{sing} we will explain why some regions of the phase space are not affected by any cosmological singularity, and in Sect. \ref{generic} we will summarize the patterns that have emerged in our analysis by discussing which cosmological features we have discovered are sensitive to the particular modeling of the fluid, and which instead  seem to be a general property. We will conclude in Sect. \ref{conclusion} by discussing the cosmological relevance of our analysis and by putting the present work in the perspective of possible future projects.   In Appendix \ref{app:found} we review the applicability of the fluid models considered in this paper for the description of real gases beyond the cosmological context.     The analytical computations of the stability of the isolated fixed points and of the invariant submanifolds are reported in the Appendices \ref{app:stab_fin}, \ref{app:cma}, \ref{app:stab_sub}, \ref{app:stab_inf} which make use of both the standard notion of linear stability and of a much more advanced technique like the  \lq\lq center manifold analysis".

\section{Basic  equations of quadratic gravity } \label{secII}

The action of quadratic gravity  in the Ricci scalar\footnote{ In this paper we will restrict ourselves to a modified gravity model quadratic in the curvature. However, other types of corrections have been proposed in the literature, either quadratic or beyond it,  as in $f(T)$ theories with torsion \cite{revmm1}, $f(Q)$ with non-metricity \cite{revmm2}, or $f(G)$ with a Gauss-Bonnet term \cite{revmm3}.} reads as \cite{Starobinsky:1980te}
\begin{equation}\label{action}
    S=\frac{1}{2\kappa}\int d^4x\sqrt{-g}f(R) + S_m \,,
\end{equation}
with\footnote{ In principle the most generic quadratic Lagrangian in curvature should also contain the terms $R_{\alpha\beta}^2 \equiv R_{\alpha\beta} R^{\alpha\beta} $ and $R_{\alpha\beta\gamma\delta}^2 \equiv R_{\alpha\beta\gamma\delta} R^{\alpha\beta\gamma\delta}$, which can be rewritten in terms of the Euler density $E\equiv R_{\alpha\beta\gamma\delta}^2-4R_{\alpha\beta}^2+R^2$ and the Weyl curvature invariant $C\equiv R_{\alpha\beta\gamma\delta}^2-2R_{\alpha\beta}^2+\frac{R^2}{3}$. $E$ does not contribute to the equation of motion due to the Gauss-Bonnet identity whereas $C$ vanishes for FLRW metric \cite{revmm4,revmm5}. Therefore the action \eqref{action} can be taken to be the most generic quadratic Lagrangian in terms of the curvature for a homogeneous and isotropic universe.} $f(R)=R+\zeta R^2$ and $\kappa=8\pi G$, $G$ being the Newton's gravitational constant. $\zeta$ is a positive parameter quantifying the deviation of the quadratic gravity from the general relativistic Einstein-Hilbert Lagrangian at high curvature. These contributions are supposed to play an important role in the early universe driving the inflationary dynamics but diluting at later epochs \cite{hysto1,hysto2,hysto3,hysto4,hysto5,hysto6,hysto7}.  This model constitutes a specific realization of a \emph{scalar-tensor} theory of gravity because modifications in the gravity sector can be re-interpreted in the Brans-Dicke language as a new degree of freedom associated to a propagating scalar field \cite{capoequiv}. 
Moreover, $S_m$ is the aggregate matter action responsible for all the fluid content of the Universe. In this paper we will assume the cosmic matter to be a perfect fluid (it is fully characterized by its pressure $P$ and energy density $\rho$) obeying to a nonideal equation of state (pressure and energy density are not directly proportional to each other). To be more specific, we will consider some fluid models which constitute examples of evolving dark energy and/or unification of exotic and regular matter since in this latter case the sign of the pressure can change at different cosmic epochs as a consequence of the evolution of the energy density. Thus, our model is intended to study the evolution from inflationary to dark energy epoch by involving both quadratic corrections in the curvature and some nonideal fluid.

Furthermore, in light of the Copernican principle, i.e. that the universe is homogeneous and isotropic, and considering an almost spatially flat universe, our geometrical model will be based on the spacetime
\begin{equation}
\label{metric}
ds^2=-dt^2 +a^2(t) (dx^2 +dy^2 +dz^2)\,.
\end{equation}
Defining $F:=\partial f / \partial R$, and introducing the Hubble function $H:=\dot a /a$, where an overdot denotes a derivative with respect to the cosmic time, we can write the field equations for a flat Friedman universe under the action \eqref{action}   as \cite{rev1,rev2,DeFelice:2010aj}:
\begin{subequations}
\begin{eqnarray}
\label{star_fe1}
&& 3 (1+ 2\zeta R) H^2 = \rho + \zeta \left (\frac{R^2}{2} -6H \dot R \right)\,,
\\
\label{star_fe2}
&& (1+ 2\zeta R) \dot H = \zeta (H \dot R -\ddot R) -\frac{\rho +P}{2}\,,
\end{eqnarray}
\end{subequations}
where we have adopted units such that $\kappa=1$, and the Ricci scalar is related to the Hubble function via
\beq
\label{ricci}
R=\frac{6 (\dot a^2 +a \ddot a)}{a^2 } = 6(2H^2 + \dot H)\,.
\eeq 
The field equations should be complemented by the Bianchi identity
\beq
\label{encons}
\dot \rho=-3H (\rho +P)
\eeq
which governs the energy conservation of the cosmic fluid.
 Furthermore, combining (\ref{star_fe1}) with (\ref{star_fe2}) we get
\beq
\label{combined}
2\dot{H} + 3H^2 = - \frac{1}{F}\left(P  - \frac{RF-f}{2} + \Ddot{F} + 2H\dot{F}\right)\,,
\eeq
which will be invoked in what follows for providing a transparent physical  interpretation to the various quantities governing the cosmological dynamics. 
In fact, the joint effects of the matter content and of the modifications to the gravity sector can be combined into an \emph{effective} total energy density and an \emph{effective} total pressure which read as \cite[Eq.(IV.82)]{res1}:
\begin{subequations}
\begin{eqnarray}
\label{rho_eff}
&& \rho_{\rm eff} := 3H^2  =  \frac{1}{F} \left(\rho  +         \frac{RF-f}{2} - 3H\dot{F}\right)
=\frac{1}{1+2\zeta R}\left[\rho+ \zeta \left (\frac{R^2}{2} -6H \dot R \right)\right] \,, \\
\label{P_eff}
&& P_{\rm eff} := -(2 \dot H +3H^2)= -\frac{1}{F} \left( P  - \frac{RF-f}{2} + \Ddot{F} + 2H\dot{F}\right)=-\frac{1}{1+2\zeta R}\left[P+ \zeta \left(-\frac{R^2}{2} +2\ddot R +4H \dot R \right)\right] \,. \nonumber\\
\end{eqnarray}
\end{subequations}
Along this line of thinking, one can also define an \emph{effective} equation of state parameter which encodes information about both the actual cosmic fluid and the curvature effects as
\begin{equation}
\label{w_eff}
    w_{\rm eff} := \frac{P_{\rm eff}}{\rho_{\rm eff}} = -1-\frac{2\dot{H}}{3H^2}\,.
\end{equation}
For the the description of the matter content of the universe, we find convenient to follow the approach of \cite{capo} and consider the following modelings for the equations of state of the cosmic fluid separately:
\begin{subequations}
\label{pall}
\begin{eqnarray}
\label{eosrk1}
&& P(\rho) = \frac{1- (\sqrt{2} -1) \alpha \rho}{ 1- (1-\sqrt{2}) \alpha \rho}\beta \rho\,\,\,\,\,\,\,(\text{Redlich-Kwong \cite{reos1}}), \label{Redlich}\\
&& P(\rho) = \frac{\beta \rho}{1+ \alpha \rho} \,\,\,\,\,\,\,(\text{(Modified) Berthelot \cite{reos2}}), 
\label{modified}\\
&& P(\rho) = \frac{\beta \rho e^{2(1- \alpha \rho)}}{2- \alpha \rho} \,\,\,\,\,\,\,(\text{Dietrici \cite{reos3}}).
\label{Dieterici}
\end{eqnarray}
\end{subequations}
The fluid equation of state parameter defined as $w:=P/\rho$ takes respectively the forms:
\begin{subequations}
\label{wrho}
\begin{eqnarray}
&& w(\rho) = \frac{1- (\sqrt{2} -1) \alpha \rho}{ 1- (1-\sqrt{2}) \alpha \rho}\beta\,\,\,\,\,\,\,(\text{Redlich-Kwong}), 
\\
&& w(\rho) = \frac{\beta}{1+ \alpha \rho} \,\,\,\,\,\,\,(\text{(Modified) Berthelot}), 
\\
&& w(\rho) = \frac{\beta e^{2(1- \alpha \rho)}}{2- \alpha \rho} \,\,\,\,\,\,\,(\text{Dietrici}).
\end{eqnarray}
\end{subequations}
Therefore, our class of models is based on three free parameters  ($\zeta$, $\alpha$, $\beta$). Different interplay between these free parameters will affect the existence of certain equilibrium configurations and certain types of finite-time singularities that we will classify in this paper with the purpose of constraining the values that these free parameters can assume by requiring these  configurations to be physically meaningful. The two free parameters entering the equation of state of the cosmic fluid should be interpreted as: $\alpha>0$ is the temperature at which a thermodynamic phase transition occurs within the fluid, and it sets the strength of the interactions between the fluid particles since in the limit $\alpha\to 0$ all these equations of state describe an ideal fluid for which pressure and energy density are directly proportional to each other $P\simeq \beta \rho$. This latter relation also shows the connection between $\beta$ and the adiabatic speed of sound inside the fluid. The interested reader can find a more detailed review of the thermodynamic foundation of these fluid approaches in the Appendix of \cite{capo},  and we will as well mention what the original reasons for their introduction for accounting for some features of real gases were in our Appendix \ref{app:found}. More in general, these models try to provide a founded thermodynamical description of an evolving dark energy beyond ad hoc redshift parametrizations for helping its possible direct detection in the the far future. In fact, for accounting for both the Planck and weak lensing datasets, a redshift-dependent  modeling of the dark energy equation of state parameter has been assumed in the form of $w=w_0 +w_1 (1+z)$ with $w_0$ and $w_1$ free parameters \cite[Sect.6.3]{planck}. However, in this simple framework the analysis  of the cosmic microwave background 
constraints on the distance to the last scattering surface is problematic, and therefore the refined Chevallier-Polarski-Linder parametrization $w=w_0+w_1 z/(1+z)$ has been introduced \cite{CPL1,CPL2}. The Barboza-Alcaniz $w=w_0+w_1 z(1+z)/(1+z^2)$ is another proposal which can be used in the whole redshift range $z\in[1,\infty)$ \cite{CPL3}. Although these frameworks have been useful for studying the running of the dark energy potential beyond a cosmological constant, they do not try to establish the microscopic properties of such an exotic fluid which remain mysterious, calling for a  physically deeper investigation.
Finally, the functional $w(\rho)$ can be interpreted also as an energy-dependent chameleon field \cite{amanda1,amanda2}.

\section{Qualitative analysis of the dynamics of quadratic gravity with nonideal fluids} \label{main}

In this section, we will derive the dynamical equations governing the evolution of the universe (\ref{metric}) in the gravity model (\ref{action}) including some nonideal fluids by implementing the set of dimensionless variables considered in \cite{carlonicapo,carloni}. Particular attention will be devoted to the rewriting of the equation of state parameters (\ref{wrho}) as functions of such dimensionless variables which are suited for a dynamical system analysis. Then, we will set some further constraints on the values of the free parameters of our model by requiring it to be free from instabilities. Lastly, we will list the mathematical equilibria and discuss their cosmological significance (which may provide tighter restrictions on the free parameters), possible bifurcations among them and their stability. Then, we will provide a prescription for compactifying the phase space with the purpose of investigating the dynamics at its infinity, and we will reconstruct the cosmological evolution on some invariant submanifolds also by analytically finding the equations of the phase orbits. In this section we also derive a set of relationships between the dimensionless variables employed in the dynamical system analysis and the observationally-relevant cosmographic parameters.

\subsection{Derivation of the autonomous first-order dynamical system in terms of dimensionless variables}
\label{condder}

The evolution equations to investigate in the $R+\zeta R^2$  gravity are:
\begin{subequations}
\label{general}
\begin{eqnarray} 
\label{genf}
&& 3  H^2 + 18 (6 H^2 \dot H +2 H \ddot H -\dot H^2) \zeta- \rho =0\,,
\\
&& 6 [\dddot H + (\zeta+4)  H \ddot H + (4-2\zeta) \dot H^2] -\dot H  -\frac{\rho +P}{2}  =0 \,,
\\
&&\dot \rho +3H (\rho +P)=0 \,,
\end{eqnarray}
\end{subequations}
where we have obtained the first two by plugging  (\ref{ricci}) into (\ref{star_fe1})-(\ref{star_fe2}). We can note that the first equation, which constitutes the {\it Generalized Friedman equation}, is not sensitive to the specific cosmic fluid modeling $P=P(\rho)$, unlike the other two dynamical equations. Furthermore, eq. (\ref{combined}) can be rewritten in terms of the Hubble function as:
\beq
6\zeta(2\dddot H+12 H \ddot H+9 \dot H^2)+2(54 \zeta H^2+1)\dot H+3H^2+P=0\,.
\eeq
Explicitly, for flat Friedman universes filled with the fluids (\ref{Redlich}), (\ref{modified}), (\ref{Dieterici}) evolving under the action of quadratic gravity,  we get the following set of dynamical equations, respectively:
\begin{itemize}
\item Redlich-Kwong fluid:
\begin{subequations}
\begin{eqnarray} 
\label{rk1}
&& 3 H^2 + 18 (6 H^2 \dot H +2 H \ddot H -\dot H^2) \zeta- \rho =0\,,
\\
\label{rk2}
&& 6 [\dddot H + (\zeta+4)  H \ddot H + (4-2\zeta) \dot H^2] -\dot H  +\frac{[\alpha\rho(\sqrt{2}-1)(\beta-1)-\beta-1]\rho}{2[\alpha\rho (\sqrt{2}-1)+1]}  =0 \,,
\\
\label{rk3}
&& \dot \rho +3H\rho \left(1+\frac{1- (\sqrt{2} -1) \alpha \rho}{ 1- (1-\sqrt{2}) \alpha \rho}\beta   \right)=0 \,.
\end{eqnarray}
\end{subequations}

\item (Modified) Berthelot fluid:
\begin{subequations}
\begin{eqnarray} 
\label{mb1}
&& 3 H^2 + 18 (6 H^2 \dot H +2 H \ddot H -\dot H^2) \zeta- \rho =0\,,
\\
\label{mb2}
&& 6 [\dddot H + (\zeta+4)  H \ddot H + (4-2\zeta) \dot H^2] -\dot H  -\frac{(\alpha\rho+\beta+1)\rho}{2(\alpha\rho +1)}  =0 \,,
\\
\label{mb3}
&& \dot \rho +3H\rho \left(1+\frac{\beta}{1+ \alpha \rho}   \right)=0 \,.
\end{eqnarray}
\end{subequations}

\item Dietrici fluid:
\begin{subequations}
\begin{eqnarray} 
\label{d1}
&& 3 H^2 + 18 (6 H^2 \dot H +2 H \ddot H -\dot H^2) \zeta- \rho =0\,,
\\
\label{d2}
&& 6 [\dddot H + (\zeta+4)  H \ddot H + (4-2\zeta) \dot H^2] -\dot H  +\frac{[\alpha\rho-\beta e^{2(1- \alpha \rho)}-2]\rho}{2(2-\alpha\rho)}  =0 \,,
\\
\label{d3}
&& \dot \rho +3H\rho \left(1+\frac{\beta e^{2(1- \alpha \rho)}}{2- \alpha \rho}\right)=0 \,.
\end{eqnarray}
\end{subequations}

\end{itemize}
These differential equations are third order in the Hubble function (or equivalently fourth order in the scale factor), and non-linear in both the Hubble function and the energy density. Thus, it is convenient to tackle them by adopting dynamical system techniques and searching possible equilibrium configurations for clarifying their cosmological meaning and analyzing their qualitative dynamics \cite{ham1,ham2,ham3,hart}. Following the formalism of \cite{carlonicapo,carloni},  we can recast these differential equations into a first-order autonomous dynamical system in terms of the following dimensionless variables:
\begin{subequations}
\label{dyn_def}
\begin{eqnarray}
\label{xdef}
x &:= & \frac{\dot F}{F H} \,=\, 12\zeta\frac{4 H \dot H +\ddot H}{H[1+12\zeta(2H^2+\dot H)]} \,,   
\\
\label{defy}
y &:= & \frac{R}{6 H^2} \,=\, 2+\frac{\dot H}{H^2} \, \equiv \, \frac{1-3w_{\rm eff}}{2} \,,   
\\
z &:= & \frac{f}{6 F H^2} \,=\, \frac{(2 H^2 +\dot H)[1+6\zeta (2H^2 +\dot H)]}{H^2 [1+12\zeta(2H^2+\dot H)]}\,,     
\\
\label{defom}
\Omega &:= & \frac{\rho}{3F H^2} \,=\,  \frac{\rho}{3H^2 [1+12\zeta(2H^2+\dot H)]}  \,. 
\end{eqnarray}
\end{subequations}
We introduce also the following auxiliary quantity:
\beq\label{q}
\textfrak{q}(y,z):= \frac{F}{R F'}  =\frac{1+12\zeta(2H^2+\dot H)}{12\zeta(2H^2+\dot H)}\equiv \frac{y}{2(y-z)}\,.
\eeq
It is clear from \eqref{dyn_def} that these dynamical variables are undefined when $H=0$. Therefore, this particular choice of variables pushes any possible fixed point corresponding to Minkowski solutions and bounce (a cosmological bounce is an alternative to the inflationary paradigm\footnote{It has already been shown that quadratic gravity can in fact give rise to nonsingular bouncing scenarios for $\zeta<0$ \cite{Paul:2014cxa}. In this paper, we will investigate its occurrence for $\zeta>0$.}) or turnaround scenarios to the infinity of the phase space. Taking into consideration fixed points at infinity requires a global phase space analysis (see \emph{e.g.} \cite{fourthorder,Goheer:2007wx} in the context of $f(R)$ gravity), which we will investigate separately in Sect. \ref{sectcomp}.  Also, we do not expect any moment of maximum expansion at which $\dot a=0=H$ since we are considering a flat ever-expanding universe filled with the effective fluid (\ref{w_eff}). However restricting to a domain of the full solution space consisting of only ever expanding (or ever contracting) solutions, this choice of variables is very advantageous when looking for a physical interpretation of the solutions and connecting with the cosmological observables. Therefore, the expansion normalized dynamical variables in \eqref{dyn_def} are appropriate for the consideration of this paper.

The first-order autonomous dynamical system governing the evolution of the cosmological variables (\ref{dyn_def}) is\footnote{We remark that some differences should be noted between our dynamical system and the one given in \cite[Eq.~(14)]{carloni} which follow from the different signatures between our Ricci scalar (\ref{ricci}) and  \cite[Eq.~(11)]{carloni}.}: 
\begin{subequations}\label{dyn_sys}
\begin{eqnarray}
\frac{dx}{dN} &= &  -2z-x^2+(1-y)x-(3w(\rho)+1)\Omega+2 \,,   
\\
\frac{dy}{dN} &= & y(x \textfrak{q}(y,z) +4 -2y) \,,  
\\
\frac{dz}{dN} &= & z(4-x-2y )+xy\textfrak{q}(y,z)  \,,     
\\
\frac{d \Omega}{dN} &= & \Omega (1-x-2y-3w(\rho))  \,,
\end{eqnarray}
\end{subequations}
where $\rho=\rho(x,y,z,\Omega)$, $N=\ln(a(t))$ denotes the number of $e$-folds of the universe \cite{linde}, and where we have exploited the chain rule 
\beq
 \frac{d {\mathcal X}} {dN} = \frac{d {\mathcal X}} {dt} \cdot \frac{d t} {da} \cdot  \frac{d a} {dN}=\frac{\dot {\mathcal X}}{H},
\eeq
for any generic quantity $\chi=\chi(t)$. From (\ref{dyn_def}) we can write the Hubble function, its time derivative, and the fluid energy density in terms of the dimensionless variables as:
\beq
\label{hvar} 
H^2=\frac{y-z}{6\zeta y(2z-y)}\,, \qquad \dot{H}= \frac{(y-z)(y-2)}{6\zeta y(2z-y)}\,, \qquad \rho=\frac{\Omega(y-z)}{2\zeta(2z-y)^2}\,,
\eeq
which, together with the definitions (\ref{Redlich}), (\ref{modified}), (\ref{Dieterici}), allow us to rewrite the equation of state parameters as
\begin{subequations}
\label{w_yzomega}
\begin{eqnarray}
\label{eq21a}
&& w(y,z,\Omega) = \frac{2\zeta(2z-y)^2 - (\sqrt{2}-1)\alpha\Omega(y-z)}{2\zeta (2z-y)^2 + (\sqrt{2}-1)\alpha\Omega(y-z)}\beta\,\,\,\,\,\,\,(\text{Redlich-Kwong}), 
\\
&& w(y,z,\Omega) = \frac{2\beta\zeta(2z-y)^2}{2\zeta(2z-y)^2 + \alpha\Omega(y-z)} \,\,\,\,\,\,\,(\text{(Modified) Berthelot}), 
\\
&& w(y,z,\Omega) = \frac{2\beta\zeta(2z-y)^2}{4\zeta(2z-y)^2 - \alpha\Omega(y-z)}\exp\left[2 - \frac{\alpha\Omega(y-z)}{\zeta(2z-y)^2}\right] \,\,\,\,\,\,\,(\text{Dietrici}).
\end{eqnarray}
\end{subequations}
Furthermore, the {\it Generalized Friedman equation} (\ref{genf}) is reduced to the constraint
\beq\label{constraint}
y +\Omega -z-x=1\,,
\eeq
which should be used for removing one cosmological variable from the dynamical system (\ref{dyn_sys}). We choose to eliminate $x$ for a twofold reason:  the $x$-equation is apparently the most complicated one, and  the $w(\rho)$ can be naturally expressed in terms of $(y,z,\Omega)$ as done in (\ref{w_yzomega}). Keeping in mind eq.(\ref{q}), the dynamical system  (\ref{dyn_sys}) becomes:
\begin{subequations}\label{red_dyn_sys}
\begin{eqnarray}
\frac{dy}{dN} &= & \frac{y(7y - 8z - 3y^2 + 3yz + y\Omega)}{2(y-z)} \,,   
\\
\frac{dz}{dN} &= & \frac{y^3+(\Omega-7z-1)y^2+2(4z+5-\Omega)yz-2z^2(z-\Omega+5)}{2(y-z)}  \,,
\\
\label{red_dyn_sys_c}
\frac{d\Omega}{dN} &= & \Omega(2-3w(y,z,\Omega)-3y+z-\Omega)  \,.
\end{eqnarray}
\end{subequations}
There are three physical viability conditions which should be accounted for when identifying the cosmologically relevant regions inside the full 3-dimensional $y$-$z$-$\Omega$ phase space\footnote{To the best of our knowledge this is the first time that these physical viability conditions are used to constrain the viable region of the phase space spanned by the expansion normalized variables \eqref{dyn_def}.}. They are the following:
\begin{itemize}
    \item Firstly, absence of ghost instabilities in $f(R)$ gravity requires $F(R)>0$, which implies $1+2\zeta R>0$ for our scenario \cite{gta}. From \eqref{ricci}-\eqref{hvar} we can write
    \begin{equation}
    \label{ricciyz}
        R = 6\dot{H} + 12H^2 = \frac{1}{\zeta}\left(\frac{y-z}{2z-y}\right) \,,
    \end{equation}
    so that the absence of ghost instabilities requires
    \begin{equation}\label{F}
        F = 1 + 2\zeta R = \frac{y}{2z-y}>0 \,,
    \end{equation}
  which can be satisfied for
    \begin{equation}
        0 < y < 2z \qquad \text{or} \qquad 2z < y < 0\,.
    \end{equation}
     These conditions represent two disconnected regions on the first and third quadrant of the $y$-$z$ plane bounded by the line $y=2z$ and the $z$-axis. 
    \item Secondly, absence of tachyonic instabilities for a generic $f(R)$ gravity theory requires $f''(R)>0$, which in our case simply implies $\zeta>0$ \cite{gta}. From the definition of the dynamical variables \eqref{dyn_def}, we note that
    \begin{equation}\label{yz}
        y-z = \frac{RF-f}{6FH^2} = \frac{\zeta R^2}{6FH^2} > 0 \quad \Rightarrow \quad y > z\,.
    \end{equation}
    \item Finally, the weak energy condition requires the energy density to be locally non-negative:
    \begin{equation}
        \rho\geq 0  \quad \Rightarrow \quad \Omega\geq 0.
    \end{equation}
    We can observe also that  $R$ is non-negative within the semi-infinite  $y\geq0$ region. Therefore,
\begin{equation}
    \Omega=\frac{\rho}{3FH^2}\leq\frac{\rho}{3F(\zeta\to 0) H^2}\leq 1\,,
\end{equation}
where the last equality follows from the observation that in the General Relativity limit  (which corresponds to $\zeta\rightarrow 0$ in the system \eqref{general}), one recovers the usual Friedman equation
\begin{equation}
    3H^2 = \rho.
\end{equation}
\end{itemize}
To summarize, there are two disjoint regions of the phase space which are physically relevant:
\begin{equation}
\label{sector1}
0 < z < y < 2z,\,\,\, 0 \leq \Omega \leq 1 \qquad \text{and} \qquad z < y < 0,\,\,\,0 \leq \Omega \leq 1\,. 
\end{equation}
These are two distinct semi-infinite wedge-shaped sectors above the $\Omega=0$ plane, in the first and third quadrants of the $y$-$z$ plane, respectively. The region in the first quadrant is confined between the two lines $y=z$ and $y=2z$, while the region in the third quadrant is confined between the line $y=z$ and the $z$-axis. We stress that till now we have not included the boundaries of these two regions (which are represented by equalities rather than inequalities in (\ref{sector1})), because they require a more careful treatment. The plane defined by the equality $y=z$ accounts for the General Relativity limit $R + \zeta R^2 \approx R$ (which can be expressed as $\zeta\rightarrow 0$ thanks to Eq.\eqref{yz}) in which the quadratic modification in the Lagrangian is negligible with respect to the Einstein-Hilbert contribution. It is not appropriate to consider the plane $y=z$ in the analysis that follows because the dynamical variables are undefined there and the dynamical system formulation that we are adopting becomes singular on the plane $y=z$. However, this does \emph{not} prevent the origin $(y,z)=(0,0)$ to be describe a physically meaningful configuration, as it can be appreciated from
\begin{equation}
    \lim_{y\rightarrow 0, z\rightarrow 0}\frac{dy}{dN} \,=\, \lim_{y\rightarrow 0, z\rightarrow 0}\frac{dz}{dN} \, = \, 0\,.
\end{equation}
The dynamical system \eqref{red_dyn_sys} is therefore singular everywhere on the $y$-$z$ plane except along the line $y=z=0$. The other boundary of the acceptable region in the first quadrant is the plane defined by the equality $y=2z$, while for the one in the third quadrant is the $z$-$\Omega$ plane defined by the condition $y=0$. The plane $y=2z$ corresponds to the limit $R + \zeta R^2 \approx \zeta R^2$ (which is equivalent to $\zeta\rightarrow +\infty$, as it can be seen from Eq.\eqref{F}) which occurs when the quadratic modification term in the Lagrangian becomes dominant over the Einstein-Hilbert contribution. At this stage, both the planes $y=2z$ and $y=0$ can be safely included  in the physically viable region of the phase space, which is thus given by 
\begin{equation}\label{viab}
{0<z<y\leq 2z \,\cup\, z<y\leq 0 \,\cup\, y=0=z\,,\,\,0\leq\Omega\leq1}\,. 
\end{equation}
As a consistency check, one can note from \eqref{hvar} that $H^2 >0$ in both  these regions. The dynamical system (\ref{red_dyn_sys}) admits the two invariant submanifolds $y=0$ and $\Omega=0$. An invariant submanifold divides the entire phase space into two distinct regions on its both sides. Although they can at most reach the boundary, no phase trajectory can cross the invariant submanifold leaving one region and entering the other.  Also, any orbit originating from a point on an invariant submanifold will always remain on that submanifold signifying that the reconstruction of the dynamics requires knowledge on the initial data. We note that  $R$ is always negative in the third quadrant of the $y$-$z$ plane (because $y$ is negative), and thus this region cannot contain any fixed point interpreted as a De-Sitter cosmology (or any other cosmology with negative deceleration parameter). Therefore, observational datasets suggest that the cosmological evolution should not occur in this region of the phase space.

\subsection{Qualitative dynamics: equilibria, stability, and bifurcations}
\label{qualsec}

\begin{table}
    \begin{center}
    	\begin{tabular}{|c|c|c|c|c|c|c|}
    		\hline
    		&&&&&&
    		\\
    		Cosmic fluid & Fixed point & $y_{\rm eq}$ & $z_{\rm eq}$ & $\Omega_{\rm eq}$ & $w_{\rm eff}$ & Cosmology 
    		\\
    		&&&&&& \\
    		\hline 
    		&&&&&&
    		\\
    		& $\mathcal{P}_1$ & $2$ & $1$ & $0$ & $-1$ & De-Sitter-like   
    		\\ 
    		&&&&&&
    		\\
    		& $\mathcal{P}_2$ & $2$ & $1+\Omega_{\rm eq}$ & $\frac{\left(\sqrt{2}-1\right)\alpha(\beta-1)}{\left(\sqrt{2}-1\right)\alpha(\beta-1)+8(\beta+1)\zeta}$ & $-1$ & De-Sitter-like 
    		\\
    		&&&&&&
    		\\
            & $\mathcal{P}_3$ & $\frac{5+3\beta}{4}$ & $\frac{5+3\beta}{8}$ & $\frac{9}{8}(\beta-1)$ & $-\frac{1}{2}(\beta+1)$ & $\begin{cases}
                                      & \text{De-Sitter like for $\beta=1$}\\
                                      & \text{$a\sim(t_s - t)^{4/(3(1-\beta))}$ for $\beta\neq1$}
                                      \end{cases}$
    		\\
    		&&&&&&
    		\\
    		Redlich-Kwong & $\mathcal{P}_4$ & $0$ & $-5$ & $0$ & $\frac{1}{3}$ & Unphysical 
    		\\
    		&&&&&&
    		\\
    		& $\mathcal{P}_5$ & $0$ & $\Omega_{\rm eq}-5$ & $\frac{40\zeta (1+\beta)}{(\sqrt{2}-1)\alpha(\beta-1)+8\zeta (1+\beta)}$ & $\frac{1}{3}$ & Unphysical 
    		\\
    		&&&&&&
    		\\
    		& $\mathcal{P}_6$ & $0$ & $0$ & $0$ & $\frac{1}{3}$ & $a \sim t^{1/2}$ 
    		\\
    		&&&&&&
    		\\
    		& $\mathcal{P}_7$ & $0$ & $0$ & $2+3\beta$ & $\frac{1}{3}$ & $a \sim t^{1/2}$ 
    		\\
    		&&&&&& \\
    		\hline
    		&&&&&&
    		\\
    		& $\mathcal{P}_1$ & $2$ & $1$ & $0$ & $-1$ & De-Sitter-like 
    		\\ 
    		&&&&&&
    		\\
    		& $\mathcal{P}_2$ & $2$ & $1+\Omega_{\rm eq}$ & $\frac{\alpha}{\alpha-8(\beta+1)\zeta}$ & $-1$ & De-Sitter-like
    		\\
    		&&&&&&
    		\\
            & $\mathcal{P}_3$ & $\frac{5}{4}$ & $\frac{5}{8}$ & $-\frac{9}{8}$ & $-\frac{1}{2}$ & Unphysical 
    		\\
    		&&&&&&
    		\\
    		(Modified) Berthelot & $\mathcal{P}_4$ & $0$ & $-5$ & $0$ & $\frac{1}{3}$ & Unphysical 
    		\\
    		&&&&&&
    		\\
    		& $\mathcal{P}_5$ & $0$ & $\Omega_{\rm eq}-5$ & $\frac{40 \zeta(1+\beta)}{8\zeta (1+\beta)-\alpha}$ & $\frac{1}{3}$ & Unphysical 
    		\\
    		&&&&&&
    		\\
    		& $\mathcal{P}_6$ & $0$ & $0$ & $0$ & $\frac{1}{3}$ & $a \sim t^{1/2}$ 
    		\\
    		&&&&&&
    		\\
    		& $\mathcal{P}_7$ & $0$ & $0$ & $2$ & $\frac{1}{3}$ & Unphysical 
    		\\
    		&&&&&& \\
    		\hline
    		&&&&&&
    		\\
    		& $\mathcal{P}_1$ & $2$ & $1$ & $0$ & $-1$ & De-Sitter-like 
    		\\ 
    		&&&&&&
    		\\
    		 & $\mathcal{P}_2$ & $2$ & $1+\Omega_{\rm eq}$ & $\frac{\alpha}{4\zeta[W(2\beta/e^2) +4]+\alpha}$ & $-1$ &  De-Sitter-like 
    		\\
    		&&&&&&
    		\\
    	Dieterici	& $\mathcal{P}_4$ & $0$ & $-5$ & $0$ & $\frac{1}{3}$ & Unphysical
    		\\
    		&&&&&&
    		\\
    		& $\mathcal{P}_5$ & $0$ & $\Omega_{\rm eq}-5$ & $\frac{20\zeta[W(2\beta/e^2) +4]}{4\zeta[W(2\beta/e^2) +4]+\alpha}$ & $\frac{1}{3}$ & Unphysical
    		\\
    		&&&&&&
    		\\
    		& $\mathcal{P}_6$ & $0$ & $0$ & $0$ & $\frac{1}{3}$ & $a\sim t^{1/2}$ 
    		\\
    		&&&&&& \\
    		\hline
            \end{tabular}
        \end{center}
    \caption{In this Table we exhibit all the equilibrium points that can be  obtained mathematically for the dynamical system (\ref{red_dyn_sys}) once Eqs.(\ref{w_yzomega}) have been implemented for the different types of cosmic fluids. The effective equation of state parameter $w_{\rm eff}$ is computed from (\ref{defy}) and it takes into account the contributions of both the actual matter content and of the curvature effects (see also (\ref{w_eff})). We refer to the main text for a detailed explanation of why some equilibria  do not represent any meaningful cosmological model.}
    	\label{table_eqpts}
    \end{table}

In Table \ref{table_eqpts} we exhibit all the equilibrium points that  arise mathematically for the dynamical system \eqref{red_dyn_sys}, along with the corresponding cosmological solution they represent, if any. 
In fact, some of the fixed points should be ignored on physical and observational grounds:
\begin{enumerate}
    \item The point $\mathcal{P}_3$ is unphysical for the scenario of a universe filled with the (Modified) Berthelot fluid because it violates the weak energy condition. Should we consider the Redlich-Kwong fluid, it can carry a physical interpretation for $1\leq \beta \leq \frac{17}{9}$. Interestingly, the former point corresponds to the $\beta=0$ case of the latter.
    \item Any orbit that approaches the point $\mathcal{P}_4$ must reside inside the third quadrant of the $y$-$z$ plane in which the deceleration parameter is always positive. Therefore, this point should be ignored on observational ground.
    \item Similarly for the state $\mathcal{P}_5$: we can note that $0\leq\Omega_{\rm eq}\leq 1$ delivers a negative $z_{\rm eq}$ implying  that any orbit approaching $\mathcal{P}_5$ must reside within the third quadrant of the $y$-$z$ plane. Therefore, this point should also be ignored on observational ground.
    \item The fixed point $\mathcal{P}_7$ is unphysical for a universe filled with the (Modified) Berthelot fluid because it violates the energy condition $\Omega_{\rm eq}\leq 1$. In the Redlich-Kwong scenario it is physical for $-\frac{2}{3} \leq\beta\leq -\frac{1}{3}$.
\end{enumerate}
The conditions on the model parameters which should be imposed for endowing the remaining mathematical solutions reported in Table \ref{table_eqpts} with a cosmological interpretation are listed in Table \ref{table_ext_cond}. They follow by imposing $0\leq\Omega\leq 1$. It should be appreciated that this affects only the range of validity of $\beta$, while no further constraints other than the already discussed are arising for $\alpha$ and $\zeta$. 

Among the physically viable fixed points we can identify three distinct types of cosmological solutions:
\begin{enumerate}
\item \emph{De-Sitter-like cosmology}:  There are  two  different possible realizations of a De-Sitter-like cosmology\footnote{Here, by {\it De-Sitter-like cosmology} we mean a cosmology in which the Hubble function is constant. From the general system of equations (\ref{general}) we can note that also Minkowski can constitute an equilibrium solution when we consider the Redlich-Kwong, (Modified) Berthelot, and Dieterici fluid models; this would correspond to the particular case of $H=const.=0$ (and $\rho=P=0$). However, the dynamical variables (\ref{dyn_def}) are ill-defined for a Minkowski solution; we will address this limitation by compactifying the phase space in Sect. \ref{sectcomp}. We also remark that not all the fluid models currently adopted for a dark matter - dark energy unification are compatible with the Minkowski spacetime being an equilibrium solution, with the (Generalized) Chaplygin Gas and the Anton-Schmidt proposals being some examples; see discussion in \cite{epjc2020}.  }, which are represented by the isolated fixed points $\mathcal{P}_1$ and $\mathcal{P}_2$. The equilibrium $\mathcal{P}_1$ always constitutes a physical configuration for all the three fluids, whereas $\mathcal{P}_2$ is relevant in cosmology only imposing certain constraints on the model parameter $\beta$ as shown in Table \ref{table_ext_cond}. For all the three types of matter, the ideal fluid regime ($\alpha\rightarrow 0$) leads to a saddle-node bifurcation between $\mathcal{P}_1$ and $\mathcal{P}_2$. Furthermore, in the case of the Redlich-Kwong fluid model, also the equilibrium  $\mathcal{P}_3$ reduces to  $\mathcal{P}_1$ if we fix $\beta=1$. In this latter case a  pitchfork bifurcation is possible if we choose simultaneously $\alpha=0$ and $\beta=1$ \cite{wiggins}. 
\item \emph{Power law evolution}: There are up to two different possible realizations of the power law evolution ($a \sim t^{1/2}$), which are represented by the isolated fixed points $\mathcal{P}_6$ and $\mathcal{P}_7$. $\mathcal{P}_6$ is always physical for all the three fluids whereas  $\mathcal{P}_7$ is relevant for cosmology  only in the Redlich-Kwong scenario and restricting  $-\frac{2}{3} \leq \beta\leq -\frac{1}{3}$.
\item \emph{Big-Rip singularity}: The fixed point ${\mathcal P}_3$, which is physically well-defined only considering the Redlich-Kwong fluid, represents a big-rip singularity which is asymptotically approached  at the finite time\footnote{For computing $t_s$, note that $\frac{d (1/H)}{dt}=3(1-\beta)/4$, which provides $\frac{d \ln a}{dt}=\frac{4 H_0}{4+3(1-\beta)H_0 t}$, and that we fixed $a(t=0)=1$.}
\begin{equation}
\label{singtime}
    t_s = \frac{4}{3(\beta-1)H_0}\,.
\end{equation}
In fact, we can note that the scale factor is diverging by looking at its time evolution; the energy density is also diverging because of (\ref{hvar})  and taking into account that $y=2z\neq 0$, and this comes also with a divergence in the pressure because of the form of the equation of state (\ref{Redlich}). Therefore, all the conditions for the occurrence of a Big-Rip singularity are fulfilled \cite{Sergei,refstaro,class1,class1a}. As the limiting case of $\beta=1$ is approached, which we showed corresponds to a bifurcation with the De-Sitter-like cosmology, the time at which this singularity occurs is shifted at infinity. For $\beta \neq 1$, $t_s \sim 1/ H_0$ and the singularity time is comparable to the age of the Universe.   Keeping in mind the parameter range in Table \ref{table_ext_cond}, we can also see from Table \ref{table_eqpts} that both the effective and the matter parameters $w_{\rm eff},w<-1$ for this point, where for the latter $w=-\beta$ as from (\ref{eq21a}). Therefore this fixed point also corresponds to a phantom dominated phase at which the Redlich-Kwong fluid itself behaves like a phantom fluid.  Furthermore, the adiabatic speed of sound for the Redlich-Kwong fluid, which can be computed from (\ref{eosrk1}),
\beq
c_s^2 = \frac{\partial p}{\partial \rho}= \frac{[(2\sqrt{2} -3) \alpha \rho -1] \beta (\alpha \rho -1)}{[(\sqrt{2}-1) \alpha \rho +1]^2 }\,,
\eeq
once specified to ${\mathcal P}_3$ via (\ref{hvar}) delivers
\beq
c_s^2 = \frac{(2 \sqrt{2} -3) \beta}{ (\sqrt{2} -1)^2 }=-\beta\,,
\eeq
which is smaller than $-1$ in the range of interest of $\beta$. 
\end{enumerate}

\begin{table} [ht]
    \centering
    \begin{tabular}{|c|c|c|c|}
    \hline 
    	&&&\\[-0.7em]
        Fixed points & Redlich-Kwong & (Modified) Berthelot & Dietrici 
        \\ 
        [3pt]
		\hline
		&&&\\[-0.7em]
        $\mathcal{P}_1$ & Always exists & Always exists & Always exists
        \\
        [3pt]
		\hline
		&&&\\[-0.7em]
        $\mathcal{P}_2$ & $\beta\geq1\cup\beta<-1$ & $\beta<-1$ & $\beta>-2/e^2$
        \\
        [3pt]
		\hline
		&&&\\[-0.7em]
        $\mathcal{P}_3$ & $1\leq\beta\leq\frac{17}{9}$ & Unphysical & Does not exist 
        \\
        [3pt]
		\hline
		&&&\\[-0.7em]
        $\mathcal{P}_6$ & Always exists & Always exists & Always exists
        \\
        [3pt]
		\hline
		&&&\\[-0.7em]
        $\mathcal{P}_7$ & $-\frac{2}{3}\leq\beta\leq-\frac{1}{3}$ & Unphysical & Does not exist 
        \\
        [3pt]
        \hline
    \end{tabular}
    \caption{Taking into account that $\alpha, \, \zeta>0$, the  necessary conditions for promoting the solutions listed in Table \ref{table_eqpts} from mathematical to physical are derived demanding $0\leq\Omega\leq 1$. The limits $\alpha\rightarrow 0$ and $\zeta\rightarrow 0$ correspond to ideal fluid  and General Relativity, respectively. The  points $\mathcal{P}_4$ and $\mathcal{P}_5$ are not included in this Table because they belong to a region of the phase in which the deceleration parameter is always positive.}
    \label{table_ext_cond}
\end{table}

The stability nature of the fixed points is listed in Table \ref{table_stability} and detailed calculation is presented in Appendix \ref{app:stab_fin}. It is possible to note that under the assumption that $\alpha,\zeta >0$, only the parameter $\beta$, which is related to the adiabatic speed of sound within the fluid, affects the stability nature of the finite isolated fixed point.

\begin{table}
    \centering
    \begin{tabular}{|c|c|c|c|}
    \hline
        Points & Redlich-Kwong & (Modified) Berthelot & Dietrici 
        \\
        \hline
        $\mathcal{P}_1$ & $\begin{cases}
                          & \text{Saddle for $\beta\neq-1$}\\
                          & \text{Requires more analysis for $\beta=-1$}
                          \end{cases}$
                        & $\begin{cases}
                          & \text{Saddle for $\beta\neq-1$}\\
                          & \text{Requires more analysis for $\beta=-1$}\\
                          \end{cases}$
                        & $ \begin{cases}
                          & \text{Saddle for $\beta\neq-\frac{2}{e^2}$}\\
                          & \text{Requires more analysis for $\beta=-\frac{2}{e^2}$}
                          \end{cases}$
        \\
        \hline
        $\mathcal{P}_2$ & $\begin{cases}
                          & \text{Stable for $\beta<-1$}\\
                          & \text{Saddle for $\beta>1$}\\
                          & \text{Saddle for $\beta=1$}
                          \end{cases}$ & Stable & Stable
        \\
        \hline
        $\mathcal{P}_3$ & $\begin{cases}
                          & \text{Stable for $1<\beta\leq\frac{17}{9}$}\\
                          & \text{Saddle for $\beta=1$}\\
                          \end{cases}$ & --- & ---
        \\
        \hline
        $\mathcal{P}_6$ & $\begin{cases}
                          & \text{Unstable for $\beta<\frac{2}{3}$}\\
                          & \text{Saddle for $\beta>\frac{2}{3}$}\\
                          & \text{Requires \emph{c.m.a} for $\beta=\frac{2}{3}$}
                          \end{cases} $
                        &  $\begin{cases}
                          & \text{Unstable for $\beta<\frac{2}{3}$}\\
                          & \text{Saddle for $\beta>\frac{2}{3}$}\\
                          & \text{Requires \emph{c.m.a} for $\beta=\frac{2}{3}$}
                          \end{cases} $
                        & $\begin{cases}
                          & \text{Unstable for $\beta<\frac{4}{3e^2}$}\\
                          & \text{Saddle for $\beta>\frac{4}{3e^2}$}\\
                          & \text{Requires \emph{c.m.a} for $\beta=\frac{4}{3e^2}$}
                          \end{cases}$
        \\
        \hline
        $\mathcal{P}_7$ & $\begin{cases}
                          & \text{Unstable for $\beta=-\frac{2}{3}$}\\
                          & \text{Saddle for $\beta>-\frac{2}{3}$}
                          \end{cases}$  & --- & ---
        \\
        \hline
    \end{tabular}
    \caption{Stability nature of the finite fixed points. When investigating the stability of a fixed point it is important to keep in mind the range of $\beta$ for which the fixed point exists. The abbreviation \emph{c.m.a.} stands for \lq\lq Center Manifold Analysis".}
    \label{table_stability}
\end{table}

\subsection{Phase space analysis at infinity}
\label{sectcomp}

Compactification of an unbound phase space is necessary to search for any possible fixed point that lies at its infinity: thanks to this procedure the fixed points at infinity are mapped to the boundary of the corresponding compact phase space. In general all the dynamical variables can tend to infinity, which means the phase space of the theory can exhibit a unlimited extent in all the directions. There are different  prescriptions for  $f(R)$ cosmologies (see \emph{e.g.} \cite{Goheer:2007wx} for a generic $f(R)$ theory and \cite{fourthorder} for the particular $R+\zeta R^n$ theory) for compactifying the phase space in all the directions. However, in this Sect. we introduce a new compactification technique  which directly exploits the physical viability conditions we previously derived in \eqref{viab}. As we will show below, one can use these constraints to define some invariant submanifolds  that border the physically viable region of the phase space and then we are left with only one direction in which the phase space need to be compactified.

From a mathematical point of view, the dynamical system (\ref{red_dyn_sys}) is singular on the plane $y=z$. Since this plane is one of the boundaries of the region of the phase space we are interested in, this singularity can be regularized by introducing a new time variable $\tau$ such that
\beq\label{t_redef_1}
d\tau=\frac{dN}{y-z}\,,
\eeq
in terms of which the dynamical system can be re-written as
\begin{subequations}\label{red_dyn_sys_1}
\begin{eqnarray}
\frac{dy}{d\tau} &= & \frac{y(7y - 8z - 3y^2 + 3yz + y\Omega)}{2} \,,   \\
\frac{dz}{d\tau} &= & \frac{y^3+(\Omega-7z-1)y^2+2(4z+5-\Omega)yz-2z^2(z-\Omega+5)}{2}  \,,
\\
\frac{d\Omega}{d\tau} &= & \Omega(y-z)(2-3w(y,z,\Omega)-3y+z-\Omega)  \,.
\end{eqnarray}
\end{subequations}
Now one can write
\begin{subequations}
\begin{eqnarray}
\label{subyz}
&& \frac{d}{d\tau}(y-z) = -(y-z)\left[2y^2 - y(3z+4) + z(-\Omega +z+5)\right],\\
\label{suby2z}
&& \frac{d}{d\tau}(y-2z) = -\frac{1}{2}(y-2z)\left[5y^2 + y(\Omega-7z-9) + 2z(-\Omega+z+5)\right],
\end{eqnarray}
\end{subequations}
which show that the planes $y=z$ and $y=2z$ are invariant submanifolds as well. As discussed in Sect. \ref{condder}, these two planes are equivalent to the two limits $\zeta\rightarrow 0$ and $\zeta\rightarrow +\infty$ respectively. To the best of our knowledge this is the first time that the physical viability conditions which follow from the absence of ghost and tachyonic instabilities  are recast as invariant submanifolds on the phase space of quadratic gravity. Linear stability analysis reveals that the invariant submanifold $y=z$ is always attracting whereas the invariant submanifold $y=2z$ is attracting (repelling) for $y^2 +z^2 > 5(1-\Omega)^2$ ($y^2 +z^2 < 5(1-\Omega)^2$); detailed mathematical analysis is given in Appendix \ref{app:stab_sub}. 

Before proceeding any further, it is important to comment that the dynamical system in Eq.\eqref{red_dyn_sys_1} should \emph{not} be used to determine the fixed points, because time redefinitions like \eqref{t_redef_1} may introduce artificial solutions which are not appearing in the original dynamical system. For example, one can notice that the system in Eq.\eqref{red_dyn_sys_1}  has two lines of fixed points given by
\begin{equation}
 \mathcal{L}_1\equiv(y=0=z,\,0\leq\Omega\leq1) \qquad {\rm and} \qquad \mathcal{L}_2\equiv(y=z,\,\Omega=1)\,,   
\end{equation}
both of which do not occur in the original dynamical system \eqref{red_dyn_sys}. These fictitious fixed points are  a pure mathematical artefact due to the time redefinition \eqref{t_redef_1}. We stress that this and the following steps are purely mathematical treatments aimed towards compactifying the phase space by introducing appropriate invariant submanifolds.  All the finite fixed point analysis should be carried out \emph{before} these steps.

Along with $\Omega=0$, the physically relevant region of the phase space is therefore bounded by three invariant submanifolds. Since in this region the dynamical variable $\Omega$ is itself bounded ($0\leq\Omega\leq 1$), as demonstrated in Sect. \ref{condder}, one needs only to  compactify the radial direction in the $y$-$z$ plane. For achieving this goal we first switch to  plane polar coordinates in the $y$-$z$ plane
\begin{equation}
\label{polarcoord}
    y:=r\cos{\theta}\,,\qquad z:=r\sin{\theta}\,,
\end{equation}
subject to the restrictions
\begin{equation}
\label{domain}
    0 \leq r < \infty\,,\qquad \tan^{-1}\frac{1}{2} \leq \theta \leq \frac{\pi}{4}.
\end{equation}
The dynamical system (\ref{red_dyn_sys_1}) in terms of the $r$-$\theta$-$\Omega$ variables (\ref{polarcoord}) becomes
\begin{subequations}\label{rad_dyn_sys}
\begin{eqnarray}
&& \frac{dr}{d\tau} = r^2 \left[r \cos^4 \theta+\left( \frac{3(\Omega -1)}{2} -2r \sin\theta \right)\cos^3 \theta +\frac{(1-\Omega)\sin\theta -3r}{2}\cos^2\theta +(4r\sin\theta +5-\Omega)\cos\theta +(\Omega-5)\sin\theta-r \right]  \nonumber\\
\\
&& \frac{d\theta}{d\tau} = -2r^2 \cos^4 \theta +\frac{1-\Omega-2r\sin\theta}{2}r\cos^3\theta +\frac{5r+3(1-\Omega) \sin\theta}{2}r\cos^2\theta +(\Omega-1-r\sin\theta)r\cos\theta \,,
\\
&& \frac{d\Omega}{d\tau} = r\Omega(\cos\theta-\sin\theta)(2-3w(r,\theta,\Omega)-3r\cos\theta+r\sin\theta-\Omega)\,,
\end{eqnarray}
\end{subequations}
where the fluid equation of state parameters (\ref{w_yzomega}) entering the latter equation are given by
\begin{subequations}
\label{w_rthetaomega}
\begin{eqnarray}
&& w(r,\theta,\Omega) = \frac{2\zeta r(2\sin\theta-\cos\theta)^2 - (\sqrt{2}-1)\alpha\Omega(\cos\theta-\sin\theta)}{2\zeta r(2\sin\theta-\cos\theta)^2 + (\sqrt{2}-1)\alpha\Omega(\cos\theta-\sin\theta)}\beta\,\,\,\,\,\,\,(\text{Redlich-Kwong}) \, , 
\\
&& w(r,\theta,\Omega) = \frac{2\beta\zeta r(2\sin\theta-\cos\theta)^2}{2\zeta r(2\sin\theta-\cos\theta)^2 + \alpha\Omega(\cos\theta-\sin\theta)} \,\,\,\,\,\,\,(\text{Modified Berthelot})\, , 
\\
&& w(r,\theta,\Omega) = \frac{2\beta\zeta r(2\sin\theta-\cos\theta)^2}{4\zeta r(2\sin\theta-\cos\theta)^2 - \alpha\Omega(\cos\theta-\sin\theta)}\exp\left[2 - \frac{\alpha\Omega(\cos\theta-\sin\theta)}{\zeta r(2\sin\theta-\cos\theta)^2}\right] \,\,\,\,\,\,\,(\text{Dietrici})\, .
\end{eqnarray}
\end{subequations}
As we have previously remarked, the introduction of the artificial line of fixed points $\mathcal{L}_1\equiv(r=0)$ is clearly confirmed by inspecting the system in Eq.\eqref{rad_dyn_sys}. We should remove this fictitious fixed point by another time redefinition
\beq\label{t_redef_2}
d\tau^* = r d\tau\,,
\eeq
so that the dynamical system becomes 
\begin{subequations}\label{rad_dyn_sys_1}
\begin{eqnarray}
&& \frac{dr}{d\tau^*} = r \left[r \cos^4 \theta+\left( \frac{3(\Omega -1)}{2} -2r \sin\theta \right)\cos^3 \theta +\frac{(1-\Omega)\sin\theta -3r}{2}\cos^2\theta +(4r\sin\theta +5-\Omega)\cos\theta +(\Omega-5)\sin\theta-r \right] \nonumber\\
\\
&& \frac{d\theta}{d\tau^*} = -2r \cos^4 \theta +\frac{1-\Omega-2r\sin\theta}{2}\cos^3\theta +\frac{5r+3(1-\Omega) \sin\theta}{2}\cos^2\theta +(\Omega-1-r\sin\theta)\cos\theta \,,
\\
&& \frac{d\Omega}{d\tau^*} = \Omega(\cos\theta-\sin\theta)(2-3w(r,\theta,\Omega)-3r\cos\theta+r\sin\theta-\Omega)\,.
\end{eqnarray}
\end{subequations}
The radial direction can be compactified by introducing the new compact variable \cite{carlonicapo,comptexp,comptlif}
\begin{equation}
    \mathcal{R}:=\frac{r}{1+r}\,,
\end{equation}
so that $r=0$ coincides with $\mathcal{R}=0$ and $r=\infty$ is mapped onto $\mathcal{R}=1$. In terms of $\mathcal{R}$ the dynamical system to investigate is
\begin{subequations}
\begin{eqnarray} 
&& \frac{d\mathcal{R}}{d\tau^*} = -\frac{\mathcal{R}}{2}\Big[ -2 \mathcal{R} \cos^4\theta +[4\mathcal{R} \sin\theta+3(1-\Omega)(1-\mathcal{R})]\cos^3\theta +[3\mathcal{R}-(1-\Omega)(1-\mathcal{R})\sin\theta]\cos^2\theta \nonumber\\ && \qquad \qquad \qquad \qquad \qquad -[8\mathcal{R}\sin\theta+2(5-\Omega)(1-\mathcal{R})]\cos\theta +2(5-\Omega)(1-\mathcal{R})\sin\theta +2\mathcal{R}
\Big]\,,
\\
&& \frac{d\theta}{d\tau^*} = \frac{\cos\theta}{2(1-\mathcal{R})}\Big[-4\mathcal{R}\cos^3\theta+[(1-\Omega)(1-\mathcal{R})-2\mathcal{R}\sin\theta]\cos^2\theta +[3(1-\Omega)(1-\mathcal{R})\sin\theta+5\mathcal{R}]\cos\theta \nonumber\\
&& \qquad \qquad \qquad \qquad\qquad -2\mathcal{R}\sin\theta-2(1-\Omega)(1-\mathcal{R}) \Big]\,,
\\
\label{om2}
&& \frac{d\Omega}{d\tau^*} = \frac{\Omega(\cos\theta -\sin\theta)}{(1-\mathcal{R})} \Big[(\sin\theta-3\cos\theta)\mathcal{R} +(2-\Omega-3w(\mathcal{R},\theta,\Omega))(1-\mathcal{R})  \Big]
\,,
\end{eqnarray}
\end{subequations}
with
\begin{subequations}
\label{w_Rthetaomega}
\begin{eqnarray}
&& w(\mathcal{R},\theta,\Omega) = \frac{2\zeta \mathcal{R}(2\sin\theta-\cos\theta)^2 - (\sqrt{2}-1)\alpha\Omega(1-\mathcal{R})(\cos\theta-\sin\theta)}{2\zeta \mathcal{R}(2\sin\theta-\cos\theta)^2 + (\sqrt{2}-1)\alpha\Omega(1-\mathcal{R})(\cos\theta-\sin\theta)}\beta\,\,\,\,\,\,\,(\text{Redlich-Kwong}), 
\\
&& w(\mathcal{R},\theta,\Omega) = \frac{2\beta\zeta \mathcal{R}(2\sin\theta-\cos\theta)^2}{2\zeta \mathcal{R}(2\sin\theta-\cos\theta)^2 + \alpha\Omega(1-\mathcal{R})(\cos\theta-\sin\theta)} \,\,\,\,\,\,\,(\text{Modified Berthelot}), 
\\
&& w(\mathcal{R},\theta,\Omega) = \frac{2\beta\zeta \mathcal{R}(2\sin\theta-\cos\theta)^2}{4\zeta \mathcal{R}(2\sin\theta-\cos\theta)^2 - \alpha\Omega(1-\mathcal{R})(\cos\theta-\sin\theta)}\exp\left[2 - \frac{\alpha\Omega(1-\mathcal{R})(\cos\theta-\sin\theta)}{\zeta \mathcal{R}(2\sin\theta-\cos\theta)^2}\right] \,\,\,\,\,\,\,(\text{Dietrici}).\nonumber\\
&&
\end{eqnarray}
\end{subequations}
We can note that all the three fluid equations of state remain well-behaved at infinity, i.e. have finite limits as $\mathcal{R}\rightarrow 1$. The latter dynamical system has a pole at $\mathcal{R}=1$, i.e. is apparently singular at the boundary. This can again be eradicated by defining a new time variable $\eta$ as
\begin{equation}\label{t_redef_3}
    d\eta=\frac{d\tau^*}{1-\mathcal{R}}\,.
\end{equation}
Therefore, the  dynamical system governing the evolution of the compatified variables can be written as
\begin{subequations} \label{comp_dyn_sys}
\begin{eqnarray} 
\label{comp_dyn_sys_a}
&& \frac{d\mathcal{R}}{d\eta} = -\frac{\mathcal{R} (1-\mathcal{R})}{2}\Big[ -2 \mathcal{R} \cos^4\theta +[4\mathcal{R} \sin\theta+3(1-\Omega)(1-\mathcal{R})]\cos^3\theta +[3\mathcal{R}-(1-\Omega)(1-\mathcal{R})\sin\theta]\cos^2\theta \nonumber\\ && \qquad \qquad \qquad \qquad \qquad -[8\mathcal{R}\sin\theta+2(5-\Omega)(1-\mathcal{R})]\cos\theta +2(5-\Omega)(1-\mathcal{R})\sin\theta +2\mathcal{R}
\Big]\,,
\\
\label{comp_dyn_sys_b}
&& \frac{d\theta}{d\eta} = \frac{\cos\theta}{2}\Big[-4\mathcal{R}\cos^3\theta+[(1-\Omega)(1-\mathcal{R})-2\mathcal{R}\sin\theta]\cos^2\theta +[3(1-\Omega)(1-\mathcal{R})\sin\theta+5\mathcal{R}]\cos\theta \nonumber\\
&& \qquad \qquad \qquad \qquad\qquad -2\mathcal{R}\sin\theta-2(1-\Omega)(1-\mathcal{R}) \Big]\,,
\\
&& \frac{d\Omega}{d\eta} = \Omega (\cos\theta -\sin\theta) \Big[(\sin\theta-3\cos\theta)R +(2-\Omega-3w(\mathcal{R},\theta,\Omega))(1-\mathcal{R})  \Big]
\,.
\end{eqnarray}
\end{subequations}
Since only the $r$-direction can be infinite, all the asymptotic fixed points should correspond to $r\rightarrow\infty$ (or $\mathcal{R}\rightarrow 1$). Therefore we need to identify the fixed points in the $\mathcal{R}$-$\theta$-$\Omega$ phase space which fulfill $\mathcal{R}=1$. Setting $\mathcal{R}=1$ in \eqref{comp_dyn_sys}, we obtain
\begin{subequations}
\label{compactified}
\begin{eqnarray}
&& \frac{d\mathcal{R}}{d\eta}\bigg\vert_{\mathcal{R}\rightarrow 1} = 0\,,
\\
\label{compt1}
&& \frac{d\theta}{d\eta}\bigg\vert_{\mathcal{R}\rightarrow 1} =\frac{\cos\theta}{2}(1-\sin2\theta)(\cos\theta -2 \sin\theta)\,,
\\
\label{compo1}
&& \frac{d\Omega}{d\eta}\bigg\vert_{\mathcal{R}\rightarrow 1} = \Omega(\cos\theta-\sin\theta)(\sin\theta-3\cos\theta)\,.
\end{eqnarray}
\end{subequations}
Interestingly, the evolution at spatial infinity is not explicitly sensitive to the modeling of the cosmic fluid as it was observed in the case of $R^n$ gravity \cite{carlonicapo} because $w$ does not enter anylonger the dynamical system  (however we remind that we have used previously our particular equations of state for checking that they well behave at infinity). A further information that can be obtained from the analysis at infinity is that $\mathcal{R}\rightarrow 1$ is an invariant submanifold. To determine the cosmology corresponding to this submanifold first we note that using \eqref{hvar} one can write
\begin{subequations}
\begin{eqnarray}
\label{varinf}
&& \lim_{r\rightarrow\infty} H^2 = \lim_{r\rightarrow\infty} \frac{\cos\theta -\sin\theta}{6\zeta r \cos\theta(2\sin\theta -\cos\theta)} = 0\,,
\\
&& \lim_{r\rightarrow\infty} \dot{H} = \lim_{r\rightarrow\infty} \frac{(\cos\theta -\sin\theta)(r\cos\theta -2)}{6\zeta r\cos\theta(2\sin\theta -\cos\theta)} = \frac{1}{6\zeta}\left(\frac{\cos\theta -\sin\theta}{2\sin\theta -\cos\theta}\right)\,,
\\
&& \lim_{r\rightarrow\infty} \rho = \lim_{r\rightarrow\infty} \frac{\Omega(\cos\theta -\sin\theta)}{2r\zeta(2\sin\theta -\cos\theta)^2} = 0\,,
\end{eqnarray}
\end{subequations}
within the range $\tan^{-1}\left(\frac{1}{2}\right)\leq\theta<\frac{\pi}{4}$. $\dot{H}$ is positive at all points on this hypersurface whereas $H$, $\rho$ vanish. This is exactly the condition for a matter-less nonsingular bounce. We remark that had we not compactified the phase space, we would have not been able to discover this bounce solution in our cosmological models for the reasons discussed below eq. (\ref{q}).
Keeping in mind the range of $\theta$ given in (\ref{domain}), asymptotic dynamical analysis reveals the following features:
\begin{itemize}
    \item The asymptotic invariant submanifold accounted for by ${\mathcal R}=1$ is a repelling submanifold. Detailed calculation regarding the stability of this submanifold is presented in Appendix \ref{app:stab_sub}. Therefore the nonsingular bouncing solutions that lie on this submanifold may constitute past epochs of the universe.
    \item The point $\mathcal{P}_i\equiv(\mathcal{R},\theta,\Omega)=(1,\tan^{-1}\frac{1}{2},0)$ is an isolated fixed point at infinity. This fixed point, although represents a nonsingular bounce, does  \emph{not} necessarily need to be matter-less, as at this point $\tan\theta=\frac{1}{2}$. In fact, as was pointed out in \cite{Paul:2014cxa}, matter-less nonsingular bounce in $f(R)$ gravity requires the equation $RF(R)-f(R)=0$ to have a positive root $R_b$, which is not satisfied in case of $R + \zeta R^2$ gravity. Linear stability analysis reveals $\mathcal{P}_i$ is a saddle point. Stability calculation is presented in Appendix \ref{app:stab_inf}.
\end{itemize}

\subsection{Evolution on the $y=2z$ invariant submanifold}
\label{y=2zman}

\begin{figure}
	\begin{center}
		$
		\begin{array}{ccc}
		{\includegraphics[angle=0,scale=0.42]{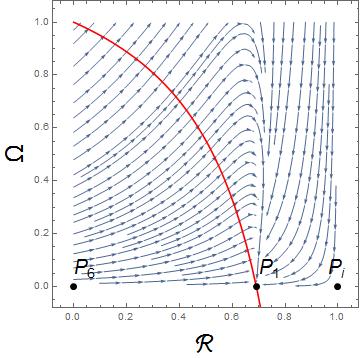}}  &
		{\includegraphics[angle=0, scale=0.42]{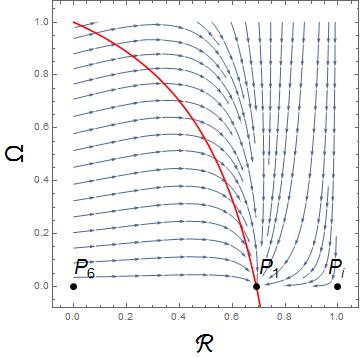}} &
		{\includegraphics[angle=0, scale=0.42]{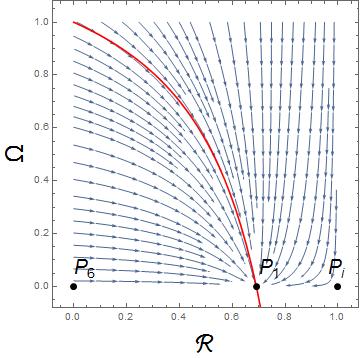}} \\ 
		(a) & (b) & (c)
		\end{array}$
	\end{center}
    \caption{Phase trajectories on the compactified $R$-$\Omega$ plane with $\theta=\tan^{-1}(1/2)$, which corresponds to the $y=2z$ submanifold, i.e. the $R^2$ regime, for (a) $w=-1$, (b) $w=0$, (c) $w=1$. In this limit the equations of state for (Modified) Berthelot and Dietrici fluids reduce to that of pressureless dust, so that they correspond to only figure (b). The equation of state for the Redlich-Kwong fluid in this limit reduces to $p=-\beta\rho$, so that this can correspond to either cases (a), (b), (c) for the parameter choice $\beta=1,0,-1$. The red curve corresponds to the boundary between the attracting part (right side of the curve) and repelling part (left side of the curve) of the submanifold. The fixed points $\mathcal{P}_1$, $\mathcal{P}_6$ and $\mathcal{P}_i$ lie on this submanifold. }
    \label{fig:y2z}
\end{figure}

The submanifold $y=2z$ corresponds to the limit $\zeta\rightarrow\infty$ which accounts for the high energy regime in which the gravitational field is so strong that the theory is dominated by the  $R^2$ term. From (\ref{suby2z}) it is seen that $y=2z$ is an invariant submanifold for the dynamics of the system. This submanifold is of attracting nature for (detailed calculations presented in Appendix \ref{app:stab_sub})
\begin{equation}
    y^2 + z^2 > 5(1-\Omega)^2 \qquad  \Leftrightarrow \qquad r > \sqrt{5}(1-\Omega) \qquad \Leftrightarrow \qquad \Omega > 1 -\frac{1}{\sqrt{5}}\left(\frac{\mathcal{R}}{1-\mathcal{R}}\right)\,,
\end{equation}
and of repelling nature for
\begin{equation}
    y^2 + z^2 < 5(1-\Omega)^2 \qquad \Leftrightarrow \qquad r < \sqrt{5}(1-\Omega) \qquad \Leftrightarrow \qquad \Omega < 1 -\frac{1}{\sqrt{5}}\left(\frac{\mathcal{R}}{1-\mathcal{R}}\right)\,.
\end{equation}
In terms of the variables $r$-$\theta$-$\Omega$ (or $\mathcal{R}$-$\theta$-$\Omega$ for the compact case) this submanifold corresponds to $\theta=\tan^{-1}(1/2)$. From (\ref{w_rthetaomega}) or (\ref{w_Rthetaomega}), for the fluid equation of state parameter we get
\begin{equation}
\lim_{\theta\rightarrow\tan^{-1}(1/2)} w = 
            {\begin{cases}
            & - \beta \quad (\text{Redlich-Kwong})\,,\\
    	    & 0 \qquad (\text{(Modified) Berthelot})\,,\\
    	    & 0 \qquad (\text{Dietrici})\,, 
    	    \end{cases}}
\end{equation}
which shows that the (Modified) Berthelot and Dieterici fluids behave like presureless dust (which may account for dark matter), and the Redlich-Kwong one behaves like an ideal fluid in which the non-linearities are suppressed.
Phase space plot on the compactified version of this submanifold plane is shown in Figure \ref{fig:y2z} for the cases of equations of state corresponding to dark matter (e.g. pressureless dust), stiff fluid and a cosmological constant. We remark that stiff fluids are canonically equivalent to massless scalar fields \cite{stiff0}, and some cosmological models indeed predict an epoch of the universe in which they are the dominating energy content \cite{stiff1,stiff2}.  

On this invariant submanifold the dynamical equations can be reduced to:
\beq
\frac{dz}{d\tau}= z [\Omega +3 (1-z)]  \,, \qquad
 \frac{d\Omega}{d\tau} = \Omega (2-\Omega -5z -3w)  \,.
\eeq
For the case of stiff matter, we can find the orbit in the phase space by solving the differential equation
\beq
\frac{d \Omega}{dz}=\frac{\Omega (1+5z +\Omega)}{z (3z -3 -\Omega)} \,,
\eeq
which delivers the implicit solution
\beq
\label{encons1}
\frac{[1+z^2+z(\Omega-2)]^2 [z^2 +2(\Omega -1)z+(1+\Omega)^2]}{[z^3+(2\Omega-3)z^2+(\Omega^2-\Omega+3)z-\Omega-1)]^2} =J_1\,,
\eeq
where $J_1$ is a constant of integration. The quantity $J_1(z, \, \Omega)$ is conserved along a particular orbit, but has different values for different orbits, and therefore it can be interpreted as the total ``energy'' of the Universe. The cosmological evolution must respect the principle of energy conservation: we can interpret eq. (\ref{encons1}) as a sort of \lq\lq energy conservation equation" which is providing a law describing how the energy of the cosmic fluid accounted for by $\Omega$ is converted into the \lq\lq geometrical energy" accounted for by the Ricci scalar $R$; this result is especially relevant for the description of the inflationary epoch in which the quadratic term in the curvature is dominating.
Furthermore, in the case of a stringy fluid with $w=-\frac{1}{3}$, which may describe some topological defects or monopoles arising in the early universe \cite{vilen}, by integrating the differential equation
\beq
\frac{d \Omega}{dz}=\frac{\Omega(5z -3 +\Omega)}{z(3z -\Omega -3)}\,,
\eeq
we obtain the implicit orbit equation
\beq
\frac{[z^3+2(\Omega-1) z^2+(\Omega^2-6\Omega+1)z+4\Omega] (z+\Omega)^2}{z [z^2+(2\Omega-1)z+\Omega^2-3 \Omega]^2}    =J_2 \,,
\eeq
where $J_2$ is a constant of integration. Also for the radiation case $w=\frac{1}{3}$ it is possible to integrate analytically the evolution equation
\beq
\frac{d \Omega}{dz}=\frac{\Omega(5z -1 +\Omega)}{z(3z -\Omega -3)}\,,
\eeq
and we obtain the implicit orbit equation
\beq
\frac{z (z-1 +\Omega)^4}{\Omega^3}  =J_3 \,,
\eeq
where $J_3$ is another constant of integration.

\subsection{Evolution on the $\Omega=0$ submanifold}
\label{omegamani}

\begin{figure}
	\begin{center}
		$
		\begin{array}{cc}
		{\includegraphics[angle=0,scale=0.45]{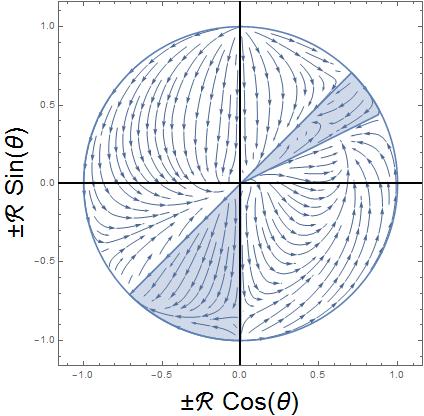}}  &
		{\includegraphics[angle=0, scale=0.45]{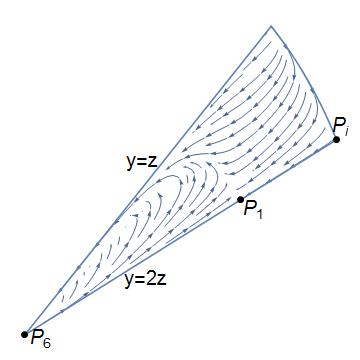}} \\ 
		(a) & (b)
		\end{array}$
	\end{center}
    \caption{In panel (a) we show the phase dynamics in the $\Omega=0$ submanifold accounted for by the evolution equations  (\ref{comp_dyn_sys_a})-(\ref{comp_dyn_sys_b}), whilst in panel (b) we focus our attention on the viable region bounded between the lines $0<z<y<2z$. This analysis provides a graphical confirmation that the submanifold ${\mathcal R}=1$ acts as a source for the dynamics, and that the cosmic evolution is indeed contained within the physical region. }
    \label{fig:pol}
\end{figure}

It appears either from (\ref{red_dyn_sys_c}) or from (\ref{om2}) that the plane $\Omega=0$ is an invariant submanifold for the cosmic dynamics. Taking into account that the physically viable region is constituted by  the wedge $0<z<y<2z$, we depict the phase orbits in this invariant submanifold  in Figure \ref{fig:pol} by using the evolution eqs. written in polar coordinates (\ref{comp_dyn_sys_a})-(\ref{comp_dyn_sys_b}). In this way we can get a graphical confirmation that the dynamics is indeed bounded inside this region and that  the boundary at spatial infinity ${\mathcal R}=1$ acts as a source for the cosmic dynamics containing possible past epochs of the universe. Unlike the case of the invariant submanifold $y=2z$ discussed in Sect. \ref{y=2zman}, the dynamics on the invariant submanifold $\Omega=0$ does not depend on the particular modeling of the cosmic fluid. However, the stability nature of this invariant submanifold is sensitive to the value of the parameter $\beta$ as demonstrated in Appendix \ref{app:stab_sub}, and more in detail it is attracting (repelling) according to $2-3\beta-3y+z<0$ ($>0$) for the Redlich-Kwong and (Modified) Berthelot fluids and $2-3e^2\beta/2-3y+z<0$ ($>0$) for the Dietrici fluid. 

\subsection{Evolution on the $\mathcal{R}=1$ submanifold}
\label{R=1sub}

$\mathcal{R}=1$ is an invariant submanifold at the infinity of the phase space. We can find the equation for the orbit $J=J(\theta, \, \Omega)$  at the infinity of the phase space by solving the partial derivative equation
    \beq
\frac{d J(\theta, \, \Omega)}{d\eta} \equiv \frac{\partial J(\theta, \, \Omega)}{\partial \theta}\frac{d\theta}{d\eta} + \frac{\partial J(\theta, \, \Omega)}{\partial \Omega}\frac{d\Omega}{d \eta}=0\,.
\eeq
Implementing (\ref{compactified}) we find 
\beq
J(\theta, \, \Omega)={\mathcal F} \left( \frac{\Omega (1-\tan \theta)^4}{(2\tan\theta-1)^5}  \right)\,,
\eeq
where ${\mathcal F}$ can be any arbitrary function. For reasons of mathematical simplicity, we choose:
\beq
J(\theta, \, \Omega)= \frac{\Omega (1-\tan \theta)^4}{(2\tan\theta-1)^5}\,.
\eeq
We note that the quantity $J(\theta, \, \Omega)$ is a positive quantity within our range of $\theta$, which is conserved along a particular orbit but can have different values for different orbits. This quantity can again be interpreted as the total ``energy'' of the Universe and the cosmological evolution must respect the principle of energy conservation. Therefore, the orbits on this submanifold are a family of curves obeying to the equation
\beq
\Omega=\frac{J (2\tan\theta-1)^5}{(1-\tan \theta)^4}\,,
\eeq
where $J$ is a constant. We stress as a consistency check that the same result  also follows by integrating a differential equation for $\frac{d\Omega}{d\theta}$ derived by dividing side by side (\ref{compo1}) with (\ref{compt1}). In terms of the original dynamical variables one can write the equation of the orbits as
\beq
\label{condomega}
\Omega=\frac{J (2z-y)^5}{y(y-z)^4}\,.
\eeq
Finally, by using (\ref{dyn_def}) this condition can be recast in terms of the energy density, of the Hubble function and of its first derivative as:
\beq
\rho( 2H^2 +\dot H)^4 -{\tilde J}H^2=0\,,
\eeq
where we have introduced the new constant
\beq
{\tilde J}=\frac{3J}{6^4 \zeta}\,.
\eeq 
This result allows us to confirm independently what written below eq. (\ref{varinf}): since $H=0$ and $\dot H \neq 0$ we get that the submanifold ${\mathcal R}=1$ corresponds to a matterless cosmological epoch. However, this should not be taken naively to imply that $\Omega=0$ because this latter quantity comes with a factor $H$ in the denominator and indeed this is true only on the hypersurface $y=2z$ as it can be understood from (\ref{condomega}).

\subsection{Cosmographic analysis }
\label{cosmop}

We will now discuss some observational properties of the universe in correspondence of the physical equilibrium points listed in Table \ref{table_eqpts} by computing the corresponding three cosmographic parameters, namely the deceleration, jerk and snap parameters \cite{visser1,Dunajski:2008tg}:
\begin{subequations}\label{cg_1}
\begin{eqnarray}
    q &\equiv& -\frac{1}{aH^2}\cdot \frac{d^2 a}{dt^2}=-1-\frac{\dot H}{H^2}\,,\\
    j &\equiv& \frac{1}{aH^3}\cdot \frac{d^3 a}{dt^2}=\frac{\ddot H}{H^3} -3q -2\,, \\
 s &\equiv& \frac{1}{aH^4}\cdot\frac{d^4 a}{dt^2}= \frac{ \dddot H}{H^4}+4j+3q(q+4)+6\,.
\end{eqnarray}
\end{subequations}
It has been shown that the cosmographic parameters are related to each other by \cite[Eq.~(15)]{jerk}, \cite[Eq.~(21)]{snap1}:
\begin{subequations}\label{cg_2}
\begin{eqnarray}
&& j = 2q^2 + q - \frac{dq}{dN}\,,
\\
&&  s = \frac{dj}{dN} - j(2 + 3q)\,.
\end{eqnarray}
\end{subequations}
The cosmographic parameters are connected to the luminosity distance via \cite{visser1,visser2,visser3,visser4,visser5}:
\beq
\label{luminosity}
d_L(z)\,\simeq \,\frac{z}{H_0} \left[1+\frac{(1-q_0)z}{2}+\frac{(-1+q_0+3 q_0^2 +j_0)z^2}{6} +\frac{(2-2q_0-15 q_0^2 -15 q_0^3 +5 j_0 +10 q_0 j_0 +s_0)z^3}{24}  \right]
\eeq
and to the cosmic history of the universe as:
\beq
\label{hcosmo}
H(z)\, \simeq \, H_0 \left[1+(1+q_0)z+\frac{(j_0 - q_0^2)z^2}{2} +\frac{(3q^2_0 + 3q^3_0
- j_0(3 + 4q_0) - s_0)z^3}{6}\right]\,,
\eeq
where a subscript `$0$' denotes that the quantity has been evaluated at the present time. In this Sect. instead we will estimate the cosmographic parameters characterizing the relevant equilibrium configurations. 
We exhibit our findings in Table \ref{tablecosmo}. We will achieve this goal by recasting the dimensionless cosmographic parameters $q$, $j$, and $s$ in terms of the dimensionless variables introduced in (\ref{dyn_def}).
Using the inter-relations between the cosmographic parameters \eqref{cg_2}, we can write 
\begin{subequations}
\begin{eqnarray}
&& q = 1-y\,,\\
&& j = 3 -5y +2y^2 +\frac{dy}{dN}\,,\\
&& s = -j(2+3q) -(5-4y)\frac{dy}{dN} +\left(\frac{dy}{dN}\right)_{,y}\frac{dy}{dN} +\left(\frac{dy}{dN}\right)_{,z}\frac{dz}{dN} +\left(\frac{dy}{dN}\right)_{,\Omega}\frac{d\Omega}{dN}\,.
\end{eqnarray}
\end{subequations}
Calculating the right hand side of the above equations using the dynamical evolution \eqref{red_dyn_sys}, we can provide explicit expressions for the cosmographic parameters in terms of the phase space coordinates:
\begin{subequations}\label{cg_3}
\begin{eqnarray}
\label{decy}
&& q = 1-y\,,\\
&& j = 3 -y +\frac{1}{2}y^2 -\frac{1}{2}\left(\frac{y^2}{y-z}\right)(1-\Omega)\,,\\
\label{sw}
&& s = -15 +10y -\frac{1}{2}y^2 -\frac{y^3}{y-z} +\frac{1}{2}\left(\frac{y^2}{y-z}\right)(3 -2\Omega -3w(y,z,\Omega)\Omega)\,.
\end{eqnarray}
\end{subequations}
These expressions can be directly generalized to include the fixed points at the infinity of the phase space by switching to the compact phase space coordinates
\begin{equation}
    y = \left(\frac{\mathcal{R}}{1-\mathcal{R}}\right)\cos\theta \,, \qquad z = \left(\frac{\mathcal{R}}{1-\mathcal{R}}\right)\sin\theta \,.
\end{equation}
Substituting in Eq.\eqref{cg_3} we get the following explicit expressions for the cosmographic parameters in terms of the compact phase space coordinates:
\begin{subequations}
\begin{eqnarray}
&& q = 1 -\left(\frac{\mathcal{R}}{1-\mathcal{R}}\right)\cos\theta\,,
\\
&& j = 3 -\left(\frac{\mathcal{R}}{1-\mathcal{R}}\right)\cos\theta +\frac{1}{2}\left(\frac{\mathcal{R}}{1-\mathcal{R}}\right)^2 \cos^2 \theta -\frac{1}{2}\left(\frac{\mathcal{R}}{1-\mathcal{R}}\right)\left(\frac{\cos^2 \theta}{\cos\theta - \sin\theta}\right)(1-\Omega)\,,
\\
&& s = -15 +10\left(\frac{\mathcal{R}}{1-\mathcal{R}}\right)\cos\theta -\frac{1}{2}\left(\frac{\mathcal{R}}{1-\mathcal{R}}\right)^2 \cos^2 \theta -\left(\frac{\mathcal{R}}{1-\mathcal{R}}\right)^2 \left(\frac{\cos^3 \theta}{\cos\theta - \sin\theta}\right) \nonumber
\\
&& \hspace{50mm} +\frac{1}{2}\left(\frac{\mathcal{R}}{1-\mathcal{R}}\right)\left(\frac{\cos^2 \theta}{\cos\theta - \sin\theta}\right)(3 -2\Omega -3w(\mathcal{R},\theta,\Omega)\Omega)\,.
\end{eqnarray}
\end{subequations}
First of all, we easily get that (\ref{decy}) implies that $y=$constant submanifolds correspond to cosmic moments with the same value of the deceleration parameter.  Possible Minkowski solutions necessarily lie on $y=1$, and therefore our models do not contain them as equilibrium configurations (this resolves the ambiguity whether the De-Sitter-like cosmologies we have identified in Sect. \ref{qualsec} can come with $H=const.=0$). The expression of the cosmographic parameters in terms of compact variables also allows us to show that the cosmographic quantities are diverging at spatial infinity of the phase space which is consistent with having a bounce there characterized by $d_L \to \infty$. We would like to mention that a cross-check procedure for computing the jerk parameter which does not rely on inter-relations is the following.
We implement (\ref{hvar}) into (\ref{xdef}) and then solve for the second time derivative of the Hubble function: 
\beq
\ddot H= \frac{(8-x)y^2-8(z+2)y+16z}{72 \zeta y (y-2z)}\sqrt{\frac{6(z-y)}{\zeta y (y-2z)}}\,,
\eeq
from which $x$ can be eliminated thanks to the constraint (\ref{constraint}):
\beq
\label{Hsec}
\ddot H= \frac{(9+z-\Omega-y)y^2-8(z+2)y+16z}{72 \zeta y (y-2z)}\sqrt{\frac{6(z-y)}{\zeta y (y-2z)}}\,.
\eeq
Finally, the jerk parameter is obtained just by algebraic manipulations. We get:
\beq
\label{jerk}
j=\frac{(x-2)y^2+2(z+3)y-6z}{2(y-z)}=\frac{y^3+(\Omega-z-3)y^2+2(z+3)y-6z}{2(y-z)}\,.
\eeq
Interestingly, the jerk parameter is regular on $y=2z$ because the divergence in $\ddot H$ has been cured by the likewise divergence in $H$. For estimating it on $y=z\neq 0$ it is appropriate to choose a different set of variables taking into account that in such case we fall back in the General Relativity framework.

The values we get for the deceleration parameter imply that phase transitions between epochs in which the expansion of the universe is accelerating and decelerating are allowed in our class of models. In particular, at least one equilibrium point comes with $q>0$ and at least two with $q<0$ for each fluid model.
Comparison between available astrophysical datasets and the predicted values of the cosmographic parameters can constrain the theory parameters of $f(R)$ theories \cite{snap1,snap2}. A cosmographic interpretation of the Gold SNeIa dataset suggests that $q_0 \simeq  - 0.90$ and $j_0 \simeq 2.7$ \cite{snap3,snap4}. It should be noted that due to the presence of $w(y,z,\Omega)$ in the expression for the cosmographic parameter $s$ (\ref{sw}), the present-day epoch would correspond to different triples $(y,z,\Omega)$ in the phase space. However, the phase space point representing today universe is located in the region $y<1$. Information on physically relevant trajectories in the phase space can therefore be obtained by noticing that from the expression of the jerk parameter in terms of the dynamical system variables (\ref{jerk}) we get
\beq
\frac{\partial j}{\partial \Omega}=\frac{y^2}{2(y-z)}\,, \qquad \frac{\partial j}{\partial z}=\frac{y^2 (\Omega -1)}{2(y-z)^2}\,,
\eeq
implying that the jerk parameter is an increasing function with respect to $\Omega$ and decreasing with respect to $z$.

\begin{table}
    \begin{center}
    	\begin{tabular}{|c|c|c|c|c|}
    		\hline
    		&&&&
    		\\
    		Cosmic fluid & Fixed point  & $q$ & $j$ & $s$ 
    		\\
    		&&&& \\
    		\hline 
    		&&&&
    		\\
 Any fluid   		&$\mathcal{P}_1$  & $-1$ & $1$ & $1$   
    		\\ 
    		&&&&
    		\\
   Any fluid   		& $\mathcal{P}_2$   & $-1$ & $1$ & $1$ 
    		\\
    		&&&&
    		\\
    Redlich-Kwong & $\mathcal{P}_3$  & $-\frac{3\beta +1}{4}$ & $\frac{9\beta^2 -1}{8}$ & $\frac{(9\beta -5)(3\beta -1)(3\beta +1)}{32}$
    		\\
    		&&&&
    		\\
   Any fluid   		& $\mathcal{P}_6$  & $1$ & $3$ & $-15$
    		\\
    		&&&&
    		\\
Redlich-Kwong		& $\mathcal{P}_7$  & $1$ & $3$ & $-15$
    		\\
    		&&&& \\
Any fluid 	& $\mathcal{P}_i$  & $\infty$ & $\infty$ & $\infty$
    		\\
    		&&&&
    		\\
    		\hline
            \end{tabular}
        \end{center}
    \caption{This Table exhibits the values of the deceleration parameter $q$, jerk parameter $j$ and snap parameter $s$ for the physically-relevant configurations listed in Table \ref{table_eqpts}. We refer to the main text on details about the mathematical steps involved in these computations. We remark that for a correct interpretation of these results it is necessary to take into account  the appropriate range of validity for the parameter $\beta$ for each equilibrium point separately, as summarized in Table \ref{table_ext_cond}. }
    	\label{tablecosmo}
    \end{table}

\section{Singularities classification} \label{sing}

In this section we will investigate the possible occurrence of finite-time singularities in the class of Friedmannian $f(R)$ cosmologies we have previously introduced for clarifying whether the different modelings of the cosmic fluid and the modifications beyond general relativity to the gravity sector affect them. In what follows we will denote with $t_s$ the time at which a singularity may occur.  Applying a literature scheme \cite{Sergei,refstaro}, we will be interested in the following five different possible types of singularity:
\begin{enumerate}
	\item {\it Big rip singularity} or Type I  is characterized by $\lim_{t \to t_s} a(t) = \infty$, $\lim_{t \to t_s} \rho_{\rm eff}(t) = \infty$, $\lim_{t \to t_s} |P_{\rm eff}(t)| = \infty$ \cite{class1,class1a};
	\item   {\it Sudden singularity} or Type II is characterized by $\lim_{t \to t_s} a(t) = a_s$, $\lim_{t \to t_s} \rho_{\rm eff}(t) = \rho_s$, $\lim_{t \to t_s} |P_{\rm eff}(t)| = \infty$ \cite{sergei4, class2,class3,class3a};
	\item   {\it Big freeze singularity} or Type III is characterized by $\lim_{t \to t_s} a(t) = a_s$, $\lim_{t \to t_s} \rho_{\rm eff}(t) = \infty$, $\lim_{t \to t_s} |P_{\rm eff}(t)| = \infty$ \cite{class4};
	\item   {\it Generalized sudden singularity} or Type IV is characterized by $\lim_{t \to t_s} a(t) = a_s$, $\lim_{t \to t_s} \rho_{\rm eff}(t) = \rho_s$, $\lim_{t \to t_s} |P_{\rm eff}(t)| = P_s$, $\lim_{t \to t_s}  H^{(i)}(t) = \infty$, $i=2,...$ \cite{sergei3, class3,class3a,class5};
	\item   {\it $w$ singularity} or Type V is characterized by $\lim_{t \to t_s} a(t) = a_s$, $\lim_{t \to t_s} \rho_{\rm eff}(t) = 0$, $\lim_{t \to t_s} |P_{\rm eff}(t)| = 0$, $\lim_{t \to t_s} w_{\rm eff} = \lim_{t \to t_s} \frac{P_{\rm eff}(t)}{\rho_{\rm eff}(t)}=\infty$ \cite{class6,class6a}.
\end{enumerate}
In this classification, we have denoted with $a_s$, $\rho_s$ and $P_s$  some finite constant values of the scale factor, the effective  energy density and its corresponding pressure at  time $t_s$. We recall that in our analysis we will assume positive $\alpha$ and $\zeta$, while we will not make any assumptions on the sign of $\beta$. We also remark that we are working with the effective values of the energy density, pressure and equation of state parameter which encode information both on the actual matter fluid and the curvature effects, as done for example in \cite{eff1,eff2,eff3,eff4}.

Before analyzing the possible occurrence of a finite-time singularity in a generic point of the phase space, we investigate the situation in correspondence of the isolated fixed points reported in Table \ref{table_eqpts}. By looking at the evolution of the scale factor, they can exhibit three different types of cosmological evolution: de Sitter-like (${\mathcal P}_1$ and ${\mathcal P}_2$ for all the three types of fluids), radiation (${\mathcal P}_6$  for all the three types of fluids, and ${\mathcal P}_7$ for Redlich-Kwong), and power-law (${\mathcal P}_3$ for Redlich-Kwong).
\begin{itemize}
    \item The de Sitter-like cosmologies do not correspond to any finite-time singularity because the  effective energy density, pressure and equation of state parameter are finite constants.
    
    \item In the case of an ``effective'' radiation domination, the scale factor ($a\sim t^{1/2}$) would approach $a_s=0$ at the time $t=0$ in correspondence of which $\rho_{\rm eff},P_{\rm eff}\sim 1/t \to \infty$, and therefore a finite-time (recalling that the present-day time is $t_0>0$) Type III singularity occurs in the past.
    
    \item The isolated fixed point ${\mathcal P}_3$ in the Redlich-Kwong scenario can correspond to a Type I singularity occurring at a finite time $t_s$ (\ref{singtime}) in future if $ 1<\beta\leq 17/9$. We note that in ${\mathcal P}_3$ 
\beq
\rho_{\rm eff}=3H^2=\frac{4}{3(\beta-1)(t_s -t)}\,,
\eeq 
which diverges also for $\beta\to 1$; however this does not imply a \emph{finite-time} singularity as can be seen from eq. (\ref{singtime}).
\end{itemize}

We will now investigate whether some type of finite-time singularity can occur in some other regions of the phase space. By using the definition of effective energy density (\ref{rho_eff}), and the relationships between the Hubble function and the dimensionless variables (\ref{hvar}), and (\ref{polarcoord}), we have
\beq
\label{rho_inf}
\rho_{\rm eff}= \frac{y-z}{2\zeta y(2z -y)} =\frac{\cos\theta -\sin\theta}{2\zeta r \cos\theta(2\sin\theta -\cos\theta)}\,.
\eeq
Furthermore, by using  eqs. (\ref{defy})-(\ref{polarcoord}) we can get the effective equation of state parameter defined in (\ref{w_eff}), and pressure  in terms of dimensionless variables as:
\begin{eqnarray}
\label{w_inf}
&& w_{\rm eff}=\frac{1-2y}{3}=\frac{1-2r \cos\theta}{3}\,, \\
\label{p_inf}
&& P_{\rm eff}=\frac{(y-z)(1-2y)}{6\zeta y(2z -y)}=\frac{(\cos\theta -\sin\theta)(1-2r\cos\theta)}{6\zeta r \cos\theta(2\sin\theta -\cos\theta)}\, .
\end{eqnarray}
First of all, we note that on the planes $y=0$ and $y=2z$, both the effective energy density (\ref{rho_inf}) and effective pressure (\ref{p_inf}) are diverging, so that two of the requirements for having either a Type I or a Type III singularity are fulfilled. We also remark that in these regions of the phase space both the Hubble function and its first derivative are diverging, as we can understand from eq. (\ref{hvar}), and therefore we have a true curvature singularity in which the Ricci scalar (\ref{ricci}) is blowing up.

More in detail, everywhere on the plane $y=0$ the effective fluid behaves like radiation, implying a Type III singularity since $a\sim t^{1/2}$ (see also the equilibrium points ${\mathcal P}_6$ in Table \ref{table_eqpts} for all the three types of fluids, and ${\mathcal P}_7$ for Redlich-Kwong); this implies also that both energy density and pressure are diverging as $\rho_{\rm eff}, P_{\rm eff} \sim H \sim t^{-1} \sim 1/a^2 \sim (1+z)^2 $ (where this latter $z$ denotes the redshift). Therefore, assuming that the present-day is at the finite-time $t_0>0$, a Type III singularity occurs in the past at the time $t=0$. 

For understanding the behavior of the singularity on the line $y=2z$, we recall that a Type I singularity would require $w_{\rm eff}<-1$ \cite{phantom}, i.e. $y>2$. Therefore the plane $y=2$ separates the line $y=2z$ into two parts on whose sides a Type I or a Type III singularity can occur; this finding is consistent with the evolution of the scale factor exhibited in Table \ref{table_eqpts}, and the previous discussion about the equilibrium point ${\mathcal P}_3 $ for the Redlich-Kwong fluid. We can provide a rough estimate of the time $t_s$ at which these singularities occur by approximating $y \approx y_s$ in a small neighborhood of the line $y=2z$ assuming that the present-time $t_0$ configuration is contained there. This implies that $\frac{d(1/H)}{dt} \approx 2-y_s$. Thus, $H(t) \approx \frac{H_0}{1+(2-y_s)(t-t_0)H_0}$ which diverges at $t_s \approx t_0 + \frac{1}{(y_s -2)H_0}$ showing that the Type I singularity would be a future singularity, while the Type III a past singularity. 


On the other hand, for having a finite energy density, but a diverging pressure we would need a diverging equation of state parameter. By looking at (\ref{w_inf}), we see that this is possible at and only at infinity, that is for $r \to \infty$. In fact, in such a regime,  by using eq. (\ref{p_inf}) we get
\beq
\label{limpeff}
\lim_{r \to \infty} |P_{\rm eff}|=\frac{\cos\theta -\sin\theta}{3 \zeta (2\sin\theta -\cos\theta)}\,,
\eeq
which can diverge if and only if $\theta=\arctan(1/2)$. Thus, a Type II singularity may occur only at the point ${\mathcal P}_i$. Moreover, in Sect. \ref{sectcomp} we have showed that $H=0$ there, i.e. we have a well-behaving de Sitter-like scale factor and a finite (zero) effective energy density fulfilling all the conditions for having a Type II singularity. We remark that should we have considered the pressure of the actual matter fluid only, a Type II singularity may have arisen in the Dieterici framework only \cite{epjc2020}. 

Moreover, by looking at the second time derivative of the Hubble function in terms of the dimensionless variables given in eq. (\ref{Hsec}), we see that a Type IV singularity {\it may} occur either along $y=0$ or along $y=2z$. This is the mildest possible singularity because it does not imply geodesic incompletness nor diverging curvature scalars.  However, in these regions of the phase space also the energy density is diverging as it can be understood from eq. (\ref{rho_inf}) violating (at least) one of the requirements in the definition of a Type IV singularity; as previously discussed also the Ricci scalar is diverging in such circumstances violating the conditions for a Type IV singularity. Interestingly, this analysis shows that the effective energy density and pressure arising from gravity modifications cannot mimic those of linearly interacting dark matter - dark energy where the latter is modeled according to the Redlich-Kwong or the (Modified) Berthelot fluid, as in those cases a type IV singularity is allowed for  certain strengths of the coupling term \cite{epjc2020}. 

Finally, by looking at (\ref{w_inf}) we see that a Type V singularity {\it may} occur only at spatial infinity for which $r \to \infty$; this would be consistent  with having also a diverging deceleration parameter there as we have found in Sect. \ref{sectcomp}. Then, by recalling (\ref{limpeff}) we see that the effective pressure can vanish if and only if $\theta=\frac{\pi}{4}$. Under these assumptions also $\rho_{\rm eff}=0$, and taking into account the discussion of Sect. \ref{R=1sub} we further have a finite scale factor fulfilling all the requirements for a Type V singularity. This result follows from the gravity modifications and  constitutes an important difference than General Relativity in which a Type V singularity has been excluded for the three types of Redlich-Kwong, (Modified) Berthelot and Dieterici fluids \cite{epjc2020}. In fact, we can observe that such type of singularity persists also in the limiting case of $\rho,P \to 0$, i.e. of absence of an actual cosmic fluid.

\section{Discussion on generic behavior}
\label{generic}

Out of the global dynamical analysis of the system that we have presented in this paper, we note that the finite fixed points $\mathcal{P}_1$, $\mathcal{P}_6$ and the asymptotic fixed point $\mathcal{P}_i$ always exist for all the three fluids irrespective of whatever values we choose for the model parameters $\alpha$, $\beta$, $\zeta$, whereas all the other fixed points either exist for a certain fluid or for a specific range of values for the model parameters, and coincide with either $\mathcal{P}_1$ or $\mathcal{P}_6$ for certain values of those model parameters. The fixed points $\mathcal{P}_1$, $\mathcal{P}_6$ and $\mathcal{P}_i$ therefore characterize some generic features of the cosmological model in quadratic gravity consisting of the three fluids under consideration. We note that all these three fixed points lie at the line of intersection of the planes $\Omega=0$ and $y=2z$. We stress that $\Omega=0$ does \emph{not} necessarily imply a vacuum solution if either $r=0$ or $y=2z$, so that these three fixed points, although lying on the $\Omega=0$ plane, should \emph{not} necessarily correspond to vacuum solutions of the $R+\zeta R^2$ gravity theory. Another point to note is that, as we had discussed before, the plane $y=2z$ corresponds to the limit $\zeta\rightarrow\infty$, so that the points lying on this plane can be interpreted to be the solutions of $f(R)=R^2$ theory of gravity. As shown in \cite{carlonicapo}, irrespective of the fluid under consideration, the phase space of $R^n$ ($n\geq2$) gravity is always 2-dimensional, which is consistent with our interpretation.

Below we explicitly point out the generic dynamical features of the scenario that we have considered.
\begin{itemize}
    \item $\mathcal{P}_1$ is a De-Sitter solution that lies on the line of intersection of the planes $\Omega=0$ and $y=2z$. This point represents the exact De-Sitter solution of $R^2$ gravity, which is the basis of Starobinski's inflationary scenario \cite{Starobinsky:1980te}. Since it is always a non-hyperbolic fixed point one needs to do a center manifold analysis to determine the stability, which is done in Appendix \ref{app:cma}. From Eq.\eqref{eigvec} we see that two of the eigenvectors of the Jacobian at that point lie \emph{on} the $\Omega=0$ plane. The eigenvector corresponding to the negative eigenvalue is along the line $(y=2z,\,\Omega=0)$, which implies that the De-Sitter solution in $R^2$ gravity is an attractor. The eigenvector corresponding to the zero eigenvalue is along the line $y+z=3$, and the center manifold analysis reveals that the dynamics is always away from the fixed point along this direction. In the complete $R+ \zeta R^2$ theory, this corresponds to an exit from the De-Sitter phase. 
    \item $\mathcal{P}_i$ is a nonsingular bouncing solution ($H=0$, $\dot{H}>0$) as discussed in Sect. \ref{sectcomp}. As demonstrated in Appendix \ref{app:stab_inf}, this point is a saddle: repelling in the direction normal to the surface $\mathcal{R}\rightarrow1$ and attracting in the directions normal to the planes $\Omega=0$ and $y=2z$. The trajectories flowing from $\mathcal{P}_i$ to $\mathcal{P}_1$ can be interpreted as early universe solutions with an inflationary phase following a nonsingular bounce\footnote{Recent research has showed that astrophysical structures, whether they exist, can survive a bounce \cite{clifton}.}. The flow at $\mathcal{P}_1$ away from it along the line $y+z=3$ in this case corresponds to the ``graceful exit''. This is consistent with the well known result that Starobinski's inflationary scenario is a transient attractor in $R + \zeta R^2$ gravity \cite{hysto7}.
    \item $\mathcal{P}_6$ is an ``effective'' radiation dominated phase ($w_{\rm eff}=\frac{1}{3}$). The trajectories flowing from  $\mathcal{P}_6$ to $\mathcal{P}_1$ can be interpreted as late time solutions with a transition from a radiation dominated epoch to a late time accelerating epoch corresponding to dark energy domination. The flow at $\mathcal{P}_1$ away from it along the line $y+z=3$ in this case implies an end to the accelerated phase of expansion, which, in GR, is possible only if the cosmological constant changes sign.
\end{itemize}
Apart from these generic features, there are some other interesting points worthwhile for explicitly commenting upon:
\begin{itemize}
    \item An interesting thing to note is that the same fixed point $\mathcal{P}_1$ can be interpreted as either an inflationary epoch or a late time acceleration epoch, depending on which of the phase trajectories we choose to consider.
    
    \item It is also worth mentioning here that we do not get any fixed point corresponding to a matter dominated epoch because we have \emph{not} considered any dust fluid that may correspond to the CDM. A matter dominated epoch requires $w_{\rm eff}=0$ or equivalently $y=\frac{1}{2}$. We note that although any trajectory flowing from $\mathcal{P}_6$ to $\mathcal{P}_1$ crosses the plane $y=\frac{1}{2}$, there is no actual fixed point  with $y=\frac{1}{2}$, and therefore no matter dominated ``phase'' in the picture.   It is however interesting that an ``effective'' radiation-like epoch is arising even without explicitly including any ultra-relativistic fluid in the picture.
    
    \item As clear from Tables \ref{table_ext_cond}-\ref{table_stability}, for specific ranges of the values of the model parameter $\beta$, the fixed points $\mathcal{P}_2$ and/or $\mathcal{P}_3$ can exist and can also be stable. In such cases there might be more than one De-Sitter phases in the complete evolution of certain cosmological solutions. The trajectories that encounter two De-Sitter fixed points ($\mathcal{P}_1$ and $\mathcal{P}_2$ \emph{or} $\mathcal{P}_3$), with $\mathcal{P}_1$ being saddle and $\mathcal{P}_2$ (\emph{or} $\mathcal{P}_3$) being stable, are particularly interesting. It should also be noted that $\mathcal{P}_2$ (\emph{or} $\mathcal{P}_3$), when exists, can only be reached after $\mathcal{P}_1$. For such solutions $\mathcal{P}_1$ can represent Starobinski's  curvature driven inflation, whereas $\mathcal{P}_2$ (\emph{or} $\mathcal{P}_3$) can represent a future attractor corresponding to the late time acceleration. 
    
    \item It is worthwhile to note that the other two model parameters, namely $\alpha$ and $\zeta$, do not affect neither the existence nor the stability nature of the fixed points, as long as they are assumed to be positive. These two parameters quantify the deviations from ideal fluid  and from GR respectively. Existence and stability of fixed points depend only on the model parameter $\beta$, which characterizes the equation of state parameter of the fluid in its ideal limit. The parameter $\alpha$ however is crucial in relation to the bifurcation of the De-Sitter fixed points. It is precisely the non-ideal nature of the fluid ($\alpha\neq0$) that makes it possible to obtain two separate De-Sitter fixed points $\mathcal{P}_1$ and $\mathcal{P}_2$, hence providing a scope for describing the early and the late time De-Sitter epochs at one go.
    
    \item The only case in which a big-rip singularity can arise in finite future is for the Redlich-Kwong fluid with $\beta>1$. In this case the De-Sitter fixed point $\mathcal{P}_2$ is a saddle, implying that the late time De-Sitter phase is an intermediate cosmological phase and \emph{not} an attractor. In this particular case the true future attractor is $\mathcal{P}_3$, which is a big-rip singularity.
\end{itemize}

The generic features and other interesting points listed in this section are the take home messages from our present study.

\section{Conclusion}
\label{conclusion}

In this paper, we have investigated some cosmological models governed by a modified Friedman and a modified Raychaudhuri equation (\ref{general}) equivalent to the following algebraic relations between the cosmographic parameters\footnote{For the relationship between the Ricci scalar and its time derivatives and the cosmographic parameters see \cite[Eq.~(15)]{cosmray}.}:
\begin{eqnarray}
\rho \,=\, 3H^2 \Omega[1 +12 \zeta H^2 (q-1) ] \,, \qquad \rho +P(\rho) \,=\, 12 H^4[6 +q^2 +8q +s+ \zeta(j -q -2q^2)] -2 H^2(q +1)\,,
\end{eqnarray}
which can be summarized into the a single expression in which the parameter $\zeta$ does not enter directly:
\beq
3\Omega (q-1) [\rho +P(\rho)] \,=\, 36 H^2\left[(q-1)(q^2+8 q+s+6)H^2+\frac{2+q-j}{12} \right]\Omega+\rho(j-q-2 q^2)  \,.
\eeq
Whether this evolution of the rate of expansion can tame some of the problems related to the Hubble tension \cite{tension1,tension2,tension3} is beyond the purpose of the present paper, but already at this stage we have demonstrated that these models come with many desirable features: they exhibit an inflationary epoch admitting a graceful exit, a radiation dominated epoch in which light elements may form \cite{peebles}, and a late-time De Sitter epoch consistent with supernovae observations \cite{sup1,sup2}. 
Furthermore, more than one De Sitter epoch in the cosmological history can  also be predicted from thermodynamical arguments \cite{pavon}. 

We have obtained these results by applying dynamical system techniques making use of both the linear stability analysis and of center manifold analysis to a Friedman universe filled with three different non-ideal fluids separately in $f(R)=R+\zeta R^2$ gravity. We have adopted a set of dimensionless variables proposed in \cite{carloni} on which we have derived the physical restrictions (\ref{viab}) for preserving the theory from ghost and tachyonic instabilities, obtaining nevertheless a model with a rich variety of cosmological behaviors as previously mentioned. It is also interesting to note that the difference between the curvature energy density and the actual matter content energy density, which can be computed from  (\ref{hvar}), reads as:
\beq
\rho_C -\rho=3H^2 -2\rho=\frac{(y-z)(2z -y -2y\Omega)}{2\zeta y (2z -y)^2}\,.
\eeq
Therefore, the two energy densities are equal on the line  $2z -y-2y \Omega=0$, which describes a configuration that can actually arise  within the physical range (\ref{viab}).  Whether this can tame some aspects of the coincidence problem \cite{coincidence} will be explored in future publications, but we should remark that this result has not required to introduce any ad hoc interaction terms between the two fluids by modifying by hands the Bianchi identities unlike in \cite{coin1,coin2,coin3}, and therefore we can appreciate already at this stage that this potential solution would not be affected by inconsistent directions of such energy flow. For example as mentioned in \cite{coincidence}, the solar system has formed at the epoch in which the abundance of dark energy  is of the same order of magnitude of the abundance of regular matter so that a local gravitational collapse can occur in a globally accelerated expanding universe, and in our picture the roles of those two fluids would be played by a gravitational effect and by an actual matter fluid separately.

We have as well derived a connection between the dynamical system variables we have adopted and the cosmographic deceleration, jerk and snap parameters. Two equilibria points ${\mathcal P}_1$ and ${\mathcal P}_2$ come with the same values of these cosmographic parameters, and while one of them (${\mathcal P}_2$) admits a well-defined energy density of the cosmic fluid, in the case of the other (${\mathcal P}_1$) it exhibits the indefinite form $0/0$. Thus, in future we will  investigate whether the same dimensionless variables used here can be connected as well to the positions of the CMBR peaks for removing this ambiguity. We have extended the dynamical system analysis up to infinity by introducing an appropriate compactification of the phase space. As far as the Redlich-Kwong, (Modified) Berthelot, and Dieterici fluids are considered, the region at infinity of the phase space does not carry only an abstract geometrical interpretation, but it corresponds to a regime in which the equation of state for the cosmic fluid reduces to $P\simeq \beta \rho$, as it can be seen from (\ref{w_rthetaomega}). Thermodynamically, this means that the interactions between the fluid constituents are suppressed as it would happen in the limit $\alpha \to 0$. This transition to the ideal behavior of $P=w(\rho)\rho$ fluids has already been met in cosmology \cite{chap8,chap9,ong}, and it has been interpreted as a form of {\it asymptotic freedom} analogue to the one which characterizes the   quark-gluon plasma \cite{gross1,gross2}, although in this case is occurring at low rather then high energy densities.

Finally, the dynamical system approach has given us the opportunity of identifying the regions of the phase space which are free from any of the known five finite-time cosmological singularity. In our cosmological models Type II and Type V singularities can occur in the past only in correspondence of the nonsingular bounce at the infinity of the phase space, the latter being a direct consequence of the modifications to the gravity sector. A Type I singularity can occur in the future along the line $y=2z$, while a Type III in the past in correspondence of the radiation dominated epochs. Our cosmological models are not affected by a Type IV singularity. Our analysis was completely classical and whether quantum gravity corrections \'a la Wheeler-DeWitt affect this picture will be clarified in a future project, as for example done in \cite{wdw}. Other interesting future projects may consist in analyzing the astrophysical data about   recombination epoch, 21-cm line excess at cosmic dawn, and Lyman $\alpha$ forest by exploiting the existence of a radiation-dominated epoch in our models;  this can tame the previously mentioned disagreement between the thermodynamical Le Chatelier-Braun principle and the fact that a dark matter epoch should have come before the dark energy one \cite{intb2,intb3} since those phenomena are usually addressed via interacting scenarios \cite{reco1,reco2,reco3}. 

\section*{Acknowledgement}
DG is a member of the GNFM working group of the Italian INDAM.

\appendix

\section{Foundation and applicability of the Redlich-Kwong, Berthelot and Dieterici fluid models}
\label{app:found}

The first attempt of  accounting for physical properties of real gases beyond their ideal behavior has been performed by the van der Waals equation of state which implements information about the finite size of the molecules and their mutual interactions assumed to be attractive at large distances and repulsive at short ones via  a Lennard-Jones type of potential. Although this proposal came with many desirable features because it can reproduce ideal gas isotherms at high temperature and it exhibits a liquid-gas coexistence phase, the experimental  collections of more and more precise data about chemical substances has called for some improved models, as for example the Redlich-Kwong, Berthelot and Dieterici formulations. These models are still based on just two free parameters which are the critical  temperature and critical pressure at the coexistence of two phases.  Van der Waals' idea of combining the two contributions for the pressure due to the volume occupied by the molecules (which sets a limit on  the fluid  compressibility),  and their internal energy (in the ideal picture molecules only have kinetic energy) simply as an algebraic sum $P=P_{\rm att.}+P_{\rm rep.}$ has been assumed also in the Berthelot and Redlich-Kwong equations of state. They have been proposed as more realistic models for accounting for datasets about the fugacity of hydrocarbons at low (close to the ambient pressure) and high pressure respectively. Intuitively  the fugacity   quantifies the fleeting properties of a material, while rigorously it is the effective  pressure of an ideal gas at the same temperature and with the same molar Gibbs free energy as the real gas; its value for a certain substance is determined from measurements of volume as a function of pressure at constant temperature. The success of the Berthelot and Redlich-Kwong formalisms is grounded in being consistent with experimental data of different substances (methane, ethane, propane, isobutane, etc...) belonging to the family of hydrocarbons just by changing the values of the two free parameters $\alpha$ and $\beta$ for each of them; before it was necessary to consider a temperature-dependent  coefficient in the second-order virial expansion to be empirically reconstructed in each case separately. Thus, this has constituted a great advantage in epochs at which computer simulations were still not widely available. The Redlich-Kwong equation of state has then been further improved by introducing a third parameter known as the acentric factor taking into account non-spherical shapes of the molecules as the Soave-Redlich-Kwong equation for a better description of nonpolar compounds \cite{soave}. For a modern treatment of such equations of state we refer to some textbooks as \cite{book1,book2}. On the other hand, the Dieterici proposal still maintains the idea that two contributions should be included in the pressure (repulsive  because molecules are assumed to be hard spheres which cannot penetrate each other, and attractive for having a bound system), but it combines them as $P=P_{\rm rep.}e^{-P_{\rm att.}}$ improving the agreement with experimental data of the compressibility factor at high pressure than the van der Waals equation  \cite{book1,book2}. In cosmology a similar way of thinking than in chemical thermodynamics has been followed by combining into a single formalism the attractive effects of regular matter and the repulsive one of dark energy: at first the van der Waals equation of state has been chosen for the cosmic fluid   \cite{waals1,waals2,waals3,waals4}, and then the Redlich-Kwong, Berthelot and Dieterici ones have been used for enlightening whether those different characteristics which have been observed in a laboratory setting come with specific signatures in cosmology  \cite{capo}.

\section{Stability analysis of finite isolated fixed points}
\label{app:stab_fin}

In this Appendix we present in some details the calculations regarding the linear stability analysis for the cosmologically relevant isolated fixed points exhibited in Table \ref{table_eqpts}. The stability nature of an isolated fixed point in the linear regime is completely determined by the eigenvalues of the Jacobian matrix evaluated at the fixed point, provided the fixed point is \emph{hyperbolic}, \emph{i.e.} none of the eigenvalues is zero. There are four distinct possibilities that may arise for a dynamical system (for the stability classification criteria see for example \cite{ham1,ham2,ham3,hart}; for the physical significance of a certain type of stability see instead \cite{coley1,coley2}):
\begin{itemize}
    \item If all the eigenvalues have positive real parts, then the fixed point is said to be \emph{unstable}. An unstable fixed point represents a past attractor in cosmology \emph{i.e.} an epoch which represents a possible initial state for a cosmological evolution.
    \item If some of the eigenvalues have positive real parts and some have negative real parts, then the fixed point is called a \emph{saddle}. A saddle fixed point represents a possible intermediate epoch for a cosmological evolution.
    \item If all the eigenvalues have negative real parts, then the fixed point is said to be \emph{stable}. A stable fixed point represents a future attractor in cosmology, \emph{i.e.} an epoch which represents a possible final state for a cosmological evolution.
    \item If two of the eigenvalues are complex conjugate to each other with vanishing real parts, then the fixed point is unstable (stable) whether the third eigenvalue is positive (negative). This represents an oscillatory approach towards the past (future) attractor. The past (future) attractor itself represents an epoch around which the cosmological solution oscillates indefinitely.
\end{itemize}
If one or more of the eigenvalues of the Jacobian matrix are zero then the fixed point is said to be \emph{non-hyperbolic}. For non-hyperbolic fixed points Jacobian eigenvalues cannot completely determine the linear stability nature, and center manifold analysis is required to determine the stability of non-hyperbolic fixed points.

In Table \ref{table_eigenvalues} we list the eigenvalues of the Jacobian matrix for the cosmologically relevant isolated fixed points presented in Table \ref{table_eqpts}. The eigenvalues are functions of the model parameters, and therefore to determine their signs one must keep in mind that $\alpha,\,\zeta>0$, and the existence conditions for the various fixed points from Table \ref{table_ext_cond}.

\begin{table}[ht]
    \centering
    \begin{tabular}{|c|c|c|c|}
    \hline
    &&& \\[-0.7em]
    
        Fixed Points & Redlich-Kwong & (Modified) Berthelot & Dietrici 
        \\
        [3 pt]
        \hline
         &&& \\[-0.7em]
        $\mathcal{P}_1$ & $-3,\, 0,\, -3(\beta+1)$ & $-3,\, 0,\, -3(\beta+1)$ & $-3, \, 0,\, -3\left(1+\frac{e^2 \beta}{2}\right)$
        \\
        [3 pt]
        \hline
         &&& \\[-0.7em]
        $\mathcal{P}_2$ & $\frac{3(\beta^2 -1)}{2\beta}$ & $-\frac{3(1 +\beta)}{\beta},$ & $-3W\left(\frac{2\beta}{e^2}\right)-\frac{12}{W\left(\frac{2\beta}{e^2}\right)}-15,$
        \\
        & $-\frac{3}{2}\left(1 \pm \sqrt{1 -\frac{2(\sqrt{2}-1)\alpha(\beta -1)}{9\zeta (\beta+1)}}\right) $ & $-\frac{3}{2}\left(1 \pm \sqrt{1 +\frac{2\alpha}{9(1+\beta)\zeta}}\right)$ & $-\frac{3}{2}\left(1 \pm \sqrt{1 -\frac{4\alpha}{9\zeta\left(W\left(\frac{2\beta }{e^2}\right)+4\right)}}\right)$
        \\
        [3 pt]
        \hline
         &&& \\[-0.7em]
        $\mathcal{P}_3$ & $-\frac{3}{2}(\beta -1)$ & --- & ---
        \\
        & $-\frac{3}{8}\left(3\beta +1 \pm \sqrt{41 -5\beta(3\beta +2)}\right)$ & --- & ---
        \\
        [3 pt]
        \hline
         &&& \\[-0.7em]
        $\mathcal{P}_6$ & $2-3\beta,4\pm\frac{1}{\sqrt{2}}$ & $2-3\beta,4\pm\frac{1}{\sqrt{2}}$ & $2-\frac{3}{2}e^2\beta,4\pm\frac{1}{\sqrt{2}}$
        \\
        [3 pt]
        \hline
         &&& \\[-0.7em]
        $\mathcal{P}_7$ & $-2-3\beta,4\pm\frac{1+3\beta}{\sqrt{2}}$ & --- & ---
        \\
        [3 pt]
        \hline
    \end{tabular}
    \caption{Eigenvalues of the Jacobian at the finite fixed points for the dynamical system \eqref{red_dyn_sys} and presented in Table \ref{table_eqpts}. We remark that the correct physical interpretation of these results require $\alpha,\,\zeta>0$, whilst the restrictions on the parameters $\beta$ can be found in Table \ref{table_ext_cond}. }
    \label{table_eigenvalues}
\end{table}

It appears that the linear stability analysis fails for the following cases:
\begin{itemize}
    \item $\mathcal{P}_1$ for all the three fluids;
    \item $\mathcal{P}_2$ with $\beta=1$ for the Redlich-Kwong fluid;
    \item $\mathcal{P}_3$ with $\beta=1$ for the Redlich-Kwong fluid;
    \item $\mathcal{P}_6$ with $\beta=\frac{2}{3}$ for the Redlich-Kwong and (Modified) Berthelot fluids, with $\beta=\frac{4}{3e^2}$ for the Dieterici fluid;
    \item $\mathcal{P}_7$ for $\beta=-\frac{2}{3}$ for the Redlich-Kwong fluid.
\end{itemize}
For case of Redlich-Kwong fluid with $\beta=-\frac{2}{3}$, we note however that the fixed point $\mathcal{P}_7$ coincides with $\mathcal{P}_6$. The fixed point $\mathcal{P}_6$ exists for all values of $\beta$, and for $\beta=-\frac{2}{3}$ it is unstable. Therefore one can conclude that the fixed point $\mathcal{P}_7$ with $\beta=-\frac{2}{3}$ for Redlich-Kwong fluid is unstable. Stability analysis in the other cases requires the application of a center manifold analysis. Also we note that for the Redlich-Kwong fluid with $\beta=1$, the fixed points $\mathcal{P}_2$ and $\mathcal{P}_3$ coincide with $\mathcal{P}_1$, implying that a center manifold analysis for $\mathcal{P}_1$ also allows us to complete the stability analysis of $\mathcal{P}_2$ and $\mathcal{P}_3$. 

\section{Center manifold analysis for $\mathcal{P}_1$}
\label{app:cma}

Center manifold analysis is significantly mathematically rigorous \cite{wiggins,carrc}, and it has been applied in cosmology  in \cite{revmd,comptexp,cma1,cma2,cma3,cma4}, just to mention a few examples.
We carry out this analysis only for the fixed point $\mathcal{P}_1\equiv(2,1,0)$, because the Jacobian at this point has a vanishing eigenvalue irrespective of the model parameters for all the three fluids. Firstly we note that in the cases of Redlich-Kwong and (Modified) Berthelot fluids with $\beta<-1$ and in the case of Dietrici fluid with $\beta<-\frac{2}{e^2}$, $\mathcal{P}_1$ is clearly a saddle and center manifold analysis is not required. In the cases of Redlich-Kwong and (Modified) Berthelot fluids with $\beta=-1$ and in the case of Dietrici fluid with $\beta=-\frac{2}{e^2}$, stability analysis of $\mathcal{P}_1$ requires beyond center manifold analysis than presented here, as two of the eigenvalues vanish, and therefore here we investigate only the cases of Redlich-Kwong and (Modified) Berthelot fluids with $\beta>-1$ and Dietrici fluid with $\beta>-\frac{2}{e^2}$. To perform a center manifold analysis we begin by shifting the fixed point to the origin by applying the coordinate translation 
\begin{equation}
    Y=y-2\,, \qquad Z=z-1\,.
\end{equation}
In terms of $Y,\,Z,\,\Omega$ the system \eqref{red_dyn_sys} becomes
\begin{subequations}\label{red_dyn_sys_2}
\begin{eqnarray}
\frac{dY}{dN} &= & \frac{(Y+2)(Y(-3Y +\Omega +3Z -2) +2\Omega -2Z)}{2(Y -Z +1)} \,,   
\\
\frac{dZ}{dN} &= & \frac{Y^3 +Y^2 (\Omega -7Z -2) +2Y\left(\Omega +4Z^2 -(\Omega +1)Z -1\right) +2\left(Z^2 +1\right)(\Omega -Z)}{2(Y -Z +1)} \,,
\\
\frac{d\Omega}{dN} &= & -\Omega(3w(Y,Z,\Omega) +3Y +\Omega -Z +3) \,,
\end{eqnarray}
\end{subequations}
with
\begin{subequations}
\begin{eqnarray}
&& w(Y,Z,\Omega) = \frac{2\zeta(Y -2Z)^2 - \left(\sqrt{2} -1\right)\alpha\Omega(Y -Z +1)}{2\zeta(Y -2Z)^2 + \left(\sqrt{2} -1\right)\alpha\Omega(Y -Z +1)}\beta \,\,\,\,\,\,\,(\text{Redlich-Kwong}), 
\\
&& w(Y,Z,\Omega) = \frac{2\beta\zeta(Y -2Z)^2}{\alpha\Omega(Y -Z +1) +2\zeta(Y -2Z)^2} \,\,\,\,\,\,\,(\text{(Modified) Berthelot}), 
\\
&& w(Y,Z,\Omega) = \frac{2\beta\zeta(Y -2Z)^2}{\alpha\Omega(-Y +Z -1) +4\zeta(Y -2 Z)^2}\exp\left[2 +\frac{\alpha\Omega(-Y +Z -1)}{\zeta(Y -2Z)^2}\right] \,\,\,\,\,\,\,(\text{Dietrici}).
\end{eqnarray}
\end{subequations}
The fixed point $\mathcal{P}_1$  corresponds to the origin in the new variables: $\mathcal{P}_1\equiv(Y,Z,\Omega)=(0,0,0)$. Jacobian at the origin corresponding to the dynamical system \eqref{red_dyn_sys_2} is:
\begin{equation}
    J(0,0,0) = \left(
\begin{array}{ccc}
 -2 & -2 & 2 \\
 -1 & -1 & 1 \\
  0 & 0 & -3({\tilde \beta} +1) \,,
\end{array}
\right)
\end{equation}
where ${\tilde \beta}=\beta$ for the Redlich-Kwong and (Modified) Berthelot fluids, while ${\tilde \beta}=e^2 \beta/2$ for the Dieterici fluid. We stress that in the ideal fluid regime $\alpha \to 0$ for which $P=w\rho$, the 33-component of the matrix would be $-3(1+w)$. The eigenvalues remain the same as given in Tab.\ref{table_eigenvalues}. The eigenvectors are 
\begin{equation}\label{eigvec}
    \left(
\begin{array}{c}
 2 \\
 1 \\
 0 
\end{array}
\right)
\,, \qquad
\left(
\begin{array}{c}
 -1 \\
  1 \\
  0 
\end{array}
\right)
\,, \qquad
\left(
\begin{array}{c}
 -\frac{2}{3{\tilde \beta}} \\
 -\frac{1}{3{\tilde \beta}} \\
  1 
\end{array}
\right)\,.
\end{equation}
The matrix that diagonalizes the Jacobian $J(0,0,0)$ is the matrix whose three columns are the three eigenvectors above:
\begin{equation}
    S = \left(
\begin{array}{ccc}
 2 & -1 & -\frac{2}{3 {\tilde \beta}} \\
 1 &  1 & -\frac{1}{3 {\tilde \beta}} \\
 0 &  0 & 1 
\end{array}
\right)\,.
\end{equation}
One can indeed verify by direct multiplication that
\begin{equation}
 S^{-1} J(0,0,0) S = \left(
\begin{array}{ccc}
 -3 & 0 &  0 \\
  0 & 0 &  0 \\
  0 & 0 & -3({\tilde \beta}+1) 
\end{array}
\right)\,.
\end{equation}
The eigenvectors of the Jacobian at a point form an orthogonal basis at that point. In the $(Y,Z,\Omega)$ coordinates the basis vectors are
\begin{equation}
    \left(
\begin{array}{c}
 1 \\
 0 \\
 0 
\end{array}
\right)
\,, \qquad
\left(
\begin{array}{c}
  0 \\
  1 \\
  0 
\end{array}
\right)
\,, \qquad
\left(
\begin{array}{c}
 0 \\
 0 \\
 1 
\end{array}
\right)\,
\end{equation}
everywhere in the $Y$-$Z$-$\Omega$ space. The particular diagonalizing matrix $S$ for the point $(Y,Z,\Omega)=(0,0,0)$ represents a coordinate transformation $(Y,Z,\Omega)\rightarrow(U,V,W)$ at that point such that the basis vectors are now along the Jacobian eigenvectors:
\begin{equation}
    \left(
\begin{array}{c}
 U \\
 V \\
 W 
\end{array}
\right) = S^{-1} \left(
\begin{array}{c}
 Y \\
 Z \\
 \Omega 
\end{array}
\right) = \left(
\begin{array}{c}
 \frac{\Omega }{3 {\tilde \beta} }+\frac{Y}{3}+\frac{Z}{3} \\
 \frac{2 Z}{3}-\frac{Y}{3} \\
 \Omega  \\
\end{array}
\right)
\end{equation}
In terms of $U,\,V,\,W$ the system \eqref{red_dyn_sys_2} becomes
\begin{subequations}\label{red_dyn_sys_3}
\begin{eqnarray}
&&\frac{dU}{d\tau}=  \frac{ {\mathcal A}+ {\mathcal B} +{\mathcal C}}{9 {\tilde \beta}^2 [-W + 3 (1 + U - 2 V) \beta]} \,,  \\
&& {\mathcal A}= -27 [3 U (1+U)^2 - 7 U (1+U) V + (2+U) V^2 + 2 V^3] {\tilde \beta}^3 + 
 W^3 ( 6 {\tilde \beta} -2) \,, \nonumber \\  
 &&{\mathcal B}= 3 W^2 {\tilde \beta} [2 + U - 7V + 3 w(U,V,W)  - 9 (1 +U -V) {\tilde \beta}] \,, \nonumber\\
&& {\mathcal C}= 9 W {\tilde \beta}^2 [4 U (1+U) + 3 V - 7 V^2 - 3 w(U,V,W) (1+U-2V) +3 ((1 + U)^2 - (1 + U) V + V^2) {\tilde \beta}] \,, \nonumber\\
&& \frac{dV}{d\tau} = -V\frac{8 W^2 + 9 [8 U (1 + U) - (4 + 23 U) V + 14 V^2] {\tilde \beta}^2 - 
  3 {\tilde \beta} W [8 + 16 U + V (-23 + 9 {\tilde \beta})]}{6 {\tilde \beta} [ 3 (1 + U - 2 V) {\tilde \beta} -W]} \,, \\
&& \frac{dW}{d\tau} =-W\left[3 +5U -4V +3w(U,V,W) +W\left(1-\frac{5}{3{\tilde\beta}}\right)\right] \,,
\end{eqnarray}
\end{subequations}
with
\begin{subequations}
\begin{eqnarray}
&& w(U,V,W) = \left(\frac{54\beta\zeta V^2 + \left(\sqrt{2}-1\right)\alpha W(W-3\beta(U-2V+1))}{54\beta\zeta V^2 - \left(\sqrt{2}-1\right)\alpha W(W-3\beta(U-2V+1))}\right)\beta \,\,\,\,\,\,\,(\text{Redlich-Kwong}), 
\\
&& w(U,V,W) = \frac{54\beta^2 \zeta V^2}{54\beta\zeta V^2 - \alpha W(W-3\beta(U-2V+1))} \,\,\,\,\,\,\,(\text{(Modified) Berthelot}), 
 \\
 && w(U,V,W) = \frac{54\beta^2 \zeta V^2}{\alpha W[W-3\beta(U-2V+1)] + 108\beta\zeta V^2}\exp\left(\frac{\alpha W[W-3\beta(U-2V+1)]}{27\beta\zeta V^2}+2\right) \,\,\,\,\,\,\,(\text{Dietrici}). \nonumber\\
\end{eqnarray}
\end{subequations}
We note that there is no linear term in $V$ in any of the equations in the system \eqref{red_dyn_sys_3}. This is because by construction the $V$-axis is along the eigenvector corresponding to the zero eigenvalue. Let us consider the phase trajectories in the neighbourhood of the fixed point $\mathcal{P}_1=(0,0,0)$. Considering only the leading contributions at the vicinity of this point, from the system \eqref{red_dyn_sys_3} we can write the following
\begin{itemize}
    \item Redlich-Kwong fluid:
    \begin{subequations}
    \begin{eqnarray}
    && \frac{dV}{dU} \approx \frac{2V}{3\beta}\left[1+(1-2\beta)\frac{3U}{2W}\right]\,,\label{vu_p1_rk}\\
    && \frac{dV}{dW} \approx -\frac{4V}{9\beta(1-\beta)}\left[1-3\beta\frac{U}{W}\right]\,,\label{vw_p1_rk}\\
    && \frac{dW}{dU} \approx -\frac{3(1-\beta)}{2}\left[ 1+\frac{3}{2\left(\frac{W}{U} -3\beta \right)} \right]\,.\label{wu_p1_rk}
    \end{eqnarray}
    \end{subequations}
    \item (Modified) Berthelot and Dietrici fluid:
    \begin{subequations}
    \begin{eqnarray}
    && \frac{dV}{dU} \approx \frac{4V}{3\tilde\beta}\left[1+3(1-\tilde\beta)\frac{U}{W}\right]\,,\label{vu_p1_mb&d}\\
    && \frac{dV}{dW} \approx -\frac{4V}{9\tilde\beta}\left[1-3\tilde\beta\frac{U}{W}\right]\,,\label{vw_p1_mb&d}\\
    && \frac{dW}{dU} \approx -3 \left[ 1+ \frac{3}{\frac{W}{U} -3 \tilde\beta} \right] \,,\label{wu_p1_mb&d}
    \end{eqnarray}
    \end{subequations}
    with $\tilde\beta=\beta,\,\frac{e^2\beta}{2}$ for (Modified) Berthelot and Dietrici fluid respectively, and where we have used $\frac{dW}{dU}\approx \frac{dV/dU}{dV/dW}$.
\end{itemize}
Keeping in mind that $\frac{W}{U}\simeq\frac{dW}{dU}$ as ${W,U}\rightarrow{0,0}$, we get from Eqs.\eqref{wu_p1_rk} and \eqref{wu_p1_mb&d} that
\begin{equation}\label{wu_p1}
    \frac{W}{U} \simeq \frac{dW}{dU} \approx
    \begin{cases}
    & \frac{3}{4} (-1+3\beta \pm \sqrt{\beta^2 +6\beta -3})\,, \qquad (\text{Redlich-Kwong})\\
    & \frac{3}{2} (-1+\tilde\beta \pm \sqrt{\tilde\beta^2 +2 \tilde\beta -3}) \,, \qquad (\text{(Modified) Berthelot and Dietrici})\,.\\
    \end{cases}
\end{equation}
Considering the leading order contribution in the vicinity of $\mathcal{P}_1\equiv(0,0,0)$, the $V-$equation from \eqref{red_dyn_sys_3} can be written as
\begin{equation}
    \frac{d\ln|V|}{d\tau} = -4\left(U -\frac{W}{3\tilde\beta}\right)\,,
\end{equation}
with $\tilde\beta=\beta$ for Redlich-Kwong, (Modified) Berthelot fluid and $\tilde\beta=\frac{e^2\beta}{2}$ for Dietrici fluid in this case. Taking one more derivative we get
\begin{equation}\label{v_p1}
    \frac{d^2\ln|V|}{d\tau^2} = -4\frac{d\ln|V|}{d\tau}\frac{d}{d\ln|V|}\left(U -\frac{W}{3\tilde\beta}\right)\,.
\end{equation}
To the leading order approximation, $\frac{dU}{d\ln|V|},\,\frac{dW}{d\ln|V|}$ are constants depending on $\beta$, whose value for different fluids can be calculated by substituting the values of $\frac{W}{U}$ from Eq.\eqref{wu_p1} into eqs.\eqref{vu_p1_rk},\eqref{vw_p1_rk},\eqref{vu_p1_mb&d},\eqref{vw_p1_mb&d}. If we define 
\begin{equation}
   \gamma = -4\frac{d}{d\ln|V|}\left(U -\frac{W}{3\tilde\beta}\right)\,,
\end{equation}
then $\gamma$ is a $\beta-$dependent constant, and the first integral of Eq.\eqref{v_p1} gives
\begin{equation}
    \frac{d\ln|V|}{d\tau} \sim e^{\gamma\tau}\,.
\end{equation}
It is clear from the above result that irrespective of the sign of $\gamma$, evolution of $V(\tau)$ is always away from the origin. The fixed point $\mathcal{P}_1$ is therefore always a saddle.
 
\section{Stability analysis of invariant submanifolds}
\label{app:stab_sub}

$X_i=\mathcal{C}$ is called an invariant submanifold of the dynamical system $\dot{\mathbf{X}}=\mathbf{f}(\mathbf{X})$ if
\begin{equation}
    \dot{X}_i \bigg\vert_{X_i=\mathcal{C}} = f_i (\mathbf{X}) \bigg\vert_{X_i=\mathcal{C}} = 0\,.
\end{equation}
Stability of an invariant submanifold is determined by the phase flow in its vicinity. If one considers a point in proximity of the submanifold with a coordinate $\mathcal{C} + \delta X_i$, then the  component of the flow  normal to the submanifold at that point is determined by
\begin{equation}
    \dot{\delta X_i} = \frac{\partial f_i}{\partial X_i}\bigg\vert_{X_i=\mathcal{C}} \delta X_i \,.
\end{equation}
If $\frac{\partial f_i}{\partial X_i}\bigg\vert_{X_i=\mathcal{C}}$ is negative (positive), then the phase flow at that point is towards (away from) the submanifold $X_i=\mathcal{C}$, and correspondingly the submanifold is attracting (repelling). If $\frac{\partial f_i}{\partial X_i}\bigg\vert_{X_i=\mathcal{C}}=0$, further analysis is required.

Armed with this concept, we can determine the stability of the invariant submanifolds that arise in our dynamical system:
\begin{itemize}
    \item The submanifold $y=2z$ can be better specified in the polar coordinate as $\theta=\tan^{-1}\left(\frac{1}{2}\right)$. From \eqref{rad_dyn_sys_1} one can compute that
    \begin{equation}
        \frac{\partial}{\partial\theta}\left(\frac{d\theta}{d\tau^*}\right)\bigg\vert_{\theta=\tan^{-1}\frac{1}{2}} = -\frac{1}{5}\left[r+\sqrt{5}(\Omega -1)\right]\,.
    \end{equation}
    Therefore the submanifold $\theta=\tan^{-1}\left(\frac{1}{2}\right)$ is attracting (repelling) for $r>\sqrt{5}(1-\Omega)$ ($r<\sqrt{5}(1-\Omega)$). In Cartesian coordinates one can state that the submanifold $y=2z$ is attracting (repelling) for $y^2 + z^2 > 5(1-\Omega)^2$ ($y^2 + z^2 < 5(1-\Omega)^2$) respectively. The line $r=\sqrt{5}(1-\Omega)$ ($y^2 + z^2 = 5(1-\Omega)^2$) separates the two regions of the submanifold with opposite dynamical characteristics.
    \item Regarding the invariant submanifold $\Omega=0$, one can compute from \eqref{red_dyn_sys_c} that
    \begin{equation}
        \frac{\partial}{\partial\Omega}\left(\frac{d\Omega}{dN}\right)\bigg\vert_{\Omega=0} = 2-3w(\Omega\rightarrow0)-3y+z\,.
    \end{equation}
    Using the expressions in Eq.\eqref{w_yzomega} to calculate $w(\Omega\rightarrow0)$, one can conclude that the invariant submanifold $\Omega=0$  is attracting (repelling) according to
    $2-3\beta-3y+z<0$ ($>0$) for the Redlich-Kwong and (Modified) Berthelot fluids and $2-3e^2\beta/2-3y+z<0$ ($>0$) for the Dietrici fluid.
    \item The submanifold $\mathcal{R}=1$ is an invariant submanifold at the infinity of the phase space. Stability of this submanifold can be determined from (\ref{comp_dyn_sys_a}) by calculating
    \begin{equation}
        \frac{\partial}{\partial{\mathcal{R}}}\left(\frac{d\mathcal{R}}{d\eta}\right)\bigg\vert_{\mathcal{R}\rightarrow1} = \frac{1}{4}(3-\cos(2\theta))(2-2\sin(2\theta)+\cos(2\theta))\,.
    \end{equation}
    The expression on the right hand side is positive within the range $\tan^{-1}\left(\frac{1}{2}\right)\leq\theta<\frac{\pi}{4}$ ($z<y\leq2z$). Therefore the invariant submanifold at infinity $\mathcal{R}=1$ is everywhere repelling.
\end{itemize}

\section{Stability analysis of fixed points at infinity}
\label{app:stab_inf}

The isolated fixed point at infinity $\mathcal{P}_i \equiv \left(1,\tan^{-1}\frac{1}{2},0\right)$ lies at the intersection of three invariant submanifolds, namely $\Omega=0$, $\theta=\tan^{-1}\left(\frac{1}{2}\right)$ and $\mathcal{R}=1$. This observation completely determines the stability nature of this fixed point. The submanifold $\mathcal{R}=1$ is everywhere repelling. The submanifold $\theta=\tan^{-1}\left(\frac{1}{2}\right)$ is attracting at $\mathcal{P}_i$ (since $\Omega=0$ and $r\rightarrow\infty$ at $\mathcal{P}_i$). The submanifold $\Omega=0$ is also attracting at $\mathcal{P}_i$ (since $-3y+z=-2y-(y-z)\rightarrow-\infty$ at $\mathcal{P}_i$, assuming $\beta$ to be finite). Therefore the fixed point $\mathcal{P}_i$ is a saddle point in the cases under consideration in this section.

{}

\end{document}